\documentclass[pra,twocolumn,showpacs,amsmath,amssymb,floatfix]
{revtex4}

\usepackage{amsmath}
\usepackage{graphicx}
\usepackage{bm}

\newcommand{\vect}[1]{\mathbf{#1}}
\newcommand{\ten}[1]{\bm{#1}}
\newcommand{\sprod}{\!\cdot\!}

\newcommand{\vprod}{\!\times\!}
\newcommand{\dif}{\mathrm{d}}
\newcommand{\mi}{i}
\newcommand{\me}{e}

\begin{document}

\title{Casimir--Polder forces on excited atoms in the strong
atom--field coupling regime}

\author{Stefan Yoshi Buhmann}
\email{s.buhmann@tpi.uni-jena.de}

\author{Dirk-Gunnar Welsch}

\affiliation{Theoretisch-Physikalisches Institut,
Friedrich-Schiller-Universit\"{a}t Jena,
Max-Wien-Platz 1, 07743 Jena, Germany}

\date{\today}

\begin{abstract}
Based on macroscopic quantum electrodynamics in linear media, we
develop a general theory of the resonant Casimir--Polder force
on an excited two-level atom in the presence of arbitrary linear
bodies, with special emphasis on the strong-coupling regime where
reabsorption of an emitted photon can give rise to (vacuum) Rabi
oscillations. We first derive a simple time-independent expression for
the force by using a dressed-state approximation. For initially
single-quantum excited atom--field systems we then study the dynamics
of the force by starting from the Lorentz force and evaluating its
average as a function of time. For strong atom--field coupling, we
find that the force may undergo damped Rabi oscillations. The damping
is due to the decay of both the atomic excitation and the field
excitation, and both amplitude and mean value of the oscillations
depend on the chosen initial state.
\end{abstract}


\maketitle


\section{Introduction}
\label{sec1}

It is well known that the presence of macroscopic bodies can
drastically change the properties of the electromagnetic field
compared to those observed in free space. A prominent manifestation of
the body-induced change of the ground-state fluctuations of the field
is the Casimir--Polder (CP) force experienced by an atom when placed
within an arrangement of unpolarized ground-state bodies (for a
review, see Ref.~\cite{0696}). Originally, CP forces were studied for
ground-state atoms where the CP potential can be identified with the
position-dependent part of the atom--field coupling energy
\cite{0030}. Since in this case the coupling involves only
off-resonant, virtual transitions of the system, the coupling energy
can be calculated by means of time-independent leading-order
perturbation theory.

For excited systems, on the contrary, real transitions must be taken
into account. In particular, when applying the method to an atom in
an excited energy eigenstate which interacts with the body-assisted
electromagnetic vacuum, one finds that the corresponding CP potential
can be significantly enhanced \cite{0042}. The enhancement is due to
the influence of now possible real transitions to lower states, with
the dependence of the force on the atomic transition frequencies
exhibiting the typical dispersion profiles in the vicinity of medium
resonances. Since these transitions are also responsible for the decay
of the atomic excitation, the application of the static approach to
the CP force on an excited atom becomes questionable.

Moreover, perturbative methods are known to break down when an atom
near-resonantly interacts with a body-assisted narrow-band field such
that the strong-coupling regime is realized; this is typically the
case when the bodies form a resonatorlike structure. For a two-level
atom that (according to the Jaynes--Cummings model \cite{0405}) is
assumed to near-resonantly interact with a single monochromatic mode,
it is well known that the energies of the exact excited energy
eigenstates of the coupled system---the so-called dressed states---are
symmetrically shifted above/below the mean of the unperturbed energy
eigenvalues. As the dressed-state energies depend on the atomic
position, they can be interpreted as CP potentials
\cite{0179,0407}---in generalization of the static approach based on
leading-order perturbation theory. In particular in the case of a
perfect standing wave in a cavity, it turns out that, depending on
the upper/lower dressed state the system is prepared in, the atom is
repelled from/attracted to the antinodes of the wave---an effect that
can strongly affect the motion of the atom \cite{0724}. Calculations
imply that atoms sufficiently slowly incident on a cavity in a
direction parallel to the standing wave can be reflected from the
cavity due to the influence of the upper dressed state
\cite{0407,0701,0418}. Here, the dependence of the transmission
probability on the velocity of the incident atoms displays
a resonance structure \cite{0720} which may be used to construct a
velocity filter for atoms \cite{0721,0725}. In the case of atoms
incident in a direction normal to the standing wave, the atoms may be
deflected during their passage through the cavity \cite{0722}.

In fact, even if a dressed state was prepared very accurately, the
system would not stay in this state forever, but it would undergo a
temporal evolution due to unavoidable dissipation, and so would the CP
force. To describe the time-dependence of CP forces acting on
excited systems, a dynamical approach is required. For weak
atom--field coupling, it has been shown that when an initially excited
atom exponentially decays to the ground state, the associated
CP force shows a similar transient behavior and eventually changes to
the ground-state force in the long-time limit \cite{0008,0012,0696}. A
two-level atom which is initially prepared in the upper state and
which strongly interacts with a (vacuum) cavity field undergoes Rabi
oscillations which are damped due to dissipation; this effect may be
phenomenologically incorporated in the Jaynes--Cummings model via
damping constants \cite{0702}. As simulated in Refs.~\cite{0705,0723},
the losses can be compensated for in a controlled way by introducing
an external pumping laser. Furthermore, by appropriate choice of the
laser intensity and frequency, it can be ensured that the system
remains in a certain steady state associated with an effective CP
potential, both the sign and magnitude of which can be continuously
controlled. Such a setup has been used to trap single (cold) atoms in
the antinodes of a standing wave in a cavity
\cite{0409,0410,0730,0726}. Since the coupling of the atom--cavity
system to the pump laser depends on the atomic position within the
standing wave, it is possible to continuously monitor the atomic
motion by recording the intensity of the pump laser leaving the cavity
\cite{0410,0726,0169,0728,0729}. This information can be used in a
feedback mechanism to enhance the trapping efficiency; whenever the
atom tends to leave the wave antinode, the trap depth is enlarged by
increasing the pump laser intensity \cite{0419}.

In the present paper, we study the resonant CP force exerted on a
two-level atom that strongly interacts with a body-assisted
electromagnetic field in more detail, where the initial state may be a
(coherent) superposition of (i) the atom being in the upper state and
the body--field system being in the ground state and (ii) the atom
being in the lower state and the body--field system being in a
single-quantum excited state. On the basis of macroscopic quantum
electrodynamics (QED) in linearly responding media, we aim at
generalizing the theory with two respects. Firstly, we allow for an
arrangement of bodies of arbitrary shape and material which are
characterized by their space- and frequency-dependent complex
permittivity and permeability, thus accounting for both material
absorption and dispersion in a natural way. The spectral and spatial
structure of the body-assisted field follows from the Green tensor of
the associated macroscopic Maxwell equations, thus generalizing the
idealized case of a single perfect standing wave in a cavity to more
realistic systems. Secondly, we include the nontrivial temporal
evolution of both the state of the system and the CP force in the
theory by developing a dynamical approach to the CP force.

The article is organized as follows. After an introduction of the
system Hamiltonian within the framework of macroscopic QED
(Sec.~\ref{sec2}), we first develop a static approach to the CP force
by assuming a dressed-state-like preparation (Sec.~\ref{sec3.1}). We
then develop a dynamical theory of the CP force (Sec.~\ref{sec3.2})
by considering the temporal evolution of the quantum averaged
Lorentz force. We end with a summary in Sec.~\ref{sec4}.


\section{Basic equations}
\label{sec2}

\subsection{Sketch of the quantization scheme}
\label{sec2.0}

Consider a neutral atom consisting of particles of charges $q_\alpha$,
masses $m_\alpha$, positions $\hat{\mathbf{r}}_\alpha$, and
canonically conjugate momenta $\hat{\mathbf{p}}_\alpha$ ($\alpha$
$\!=$ $\!1,2,\ldots$) which interact with the electromagnetic field in
the presence of linearly responding bodies. In particular, let us
consider an arrangement of magneto-electric bodies, which is described
by the spatially varying, complex permittivity
$\varepsilon(\mathbf{r},\omega)$ and permeability
$\mu(\mathbf{r},\omega)$. In electric-dipole approximation, the
multipolar-coupling Hamiltonian that governs the dynamics of the
system can be given in the form \cite{0008,0012,0696}
\begin{equation}
\label{eq1}
\hat{H}=\hat{H}_\mathrm{A}+\hat{H}_\mathrm{F}
+\hat{H}_\mathrm{AF}.
\end{equation}
Here,
\begin{equation}
\label{eq2}
\hat{H}_\mathrm{A}=\sum_{\alpha}
 \frac{\hat{\mathbf{p}}_\alpha{\!^2}}{2m_\alpha}
 +\frac{1}{2\varepsilon_0}\int\mathrm{d}^3r\,
 \hat{\mathbf{P}}_\mathrm{A}^2(\mathbf{r})
\end{equation}
is the atomic Hamiltonian, with
\begin{equation}
\label{eq3}
\hat{\mathbf{P}}_\mathrm{A}(\mathbf{r})=\sum_\alpha q_\alpha
 \hat{\bar{\mathbf{r}}}_\alpha
 \int_0^1 \mathrm{d}\lambda
 \,\delta(\mathbf{r}-\hat{\mathbf{r}}_\mathrm{A}
 -\lambda\hat{\bar{\mathbf{r}}}_\alpha)
\end{equation}
being the atomic polarization relative to the center of mass
\begin{equation}
\label{eq4}
\hat{\mathbf{r}}_\mathrm{A}=\sum_\alpha\frac{m_\alpha}{m_\mathrm{A}}
 \,\hat{\mathbf{r}}_\alpha
\end{equation}
($m_\mathrm{A}$ $\!=$ $\!\sum_\alpha m_\alpha$), where
\begin{equation}
\label{eq5}
\hat{\bar{\mathbf{r}}}_\alpha=\hat{\mathbf{r}}_\alpha
 -\hat{\mathbf{r}}_\mathrm{A}
\end{equation}
are the relative particle coordinates. The Hamiltonian
$\hat{H}_\mathrm{F}$ of the combined system of the electromagnetic
field and the magneto-electric bodies, and the atom--field
(dipo\mbox{le-)coup}ling term $\hat{H}_\mathrm{AF}$
can be expressed in terms of Bosonic (collective) variables
$\hat{\mathbf{f}}_{\lambda}(\mathbf{r},\omega)$ and
$\hat{\mathbf{f}}_{\lambda}^\dagger(\mathbf{r},\omega)$,
\begin{align}
\label{eq7}
&\bigl[\hat{\mathbf{f}}_{\lambda}(\mathbf{r},\omega),
 \hat{\mathbf{f}}_{\lambda'}^\dagger(\mathbf{r}',\omega')\bigr]
 = \delta_{\lambda\lambda'}\delta(\omega-\omega')
 \bm{\delta}(\mathbf{r}-\mathbf{r}'),\\
\label{eq8}
&\bigl[\hat{\mathbf{f}}_{\lambda}(\mathbf{r},\omega),
 \hat{\mathbf{f}}_{\lambda'}(\mathbf{r}',\omega')\bigr]=0
\end{align}
($\lambda$, $\!\lambda'$ $\!\in$ $\!\{e,m\}$), as follows:
\begin{equation}
\label{eq6}
\hat{H}_\mathrm{F}
= \sum_{\lambda=e,m}
\int\mathrm{d}^3 r \int_0^\infty \mathrm{d}\omega\,\hbar\omega\,
\hat{\mathbf{f}}_{\lambda}^\dagger(\mathbf{r},\omega)
\!\cdot\!\hat{\mathbf{f}}_{\lambda}(\mathbf{r},\omega),
\end{equation}
\begin{equation}
\label{eq9}
\hat{H}_\mathrm{AF}=-\hat{\mathbf{d}}\!\cdot\!
 \hat{\mathbf{E}}(\hat{\mathbf{r}}_\mathrm{A}),
\end{equation}
where
\begin{equation}
\label{eq10}
\hat{\mathbf{d}}
 =\sum_\alpha q_\alpha\hat{\mathbf{r}}_\alpha
 =\sum_\alpha q_\alpha\hat{\bar{\mathbf{r}}}_\alpha
\end{equation}
is the electric dipole moment of the atom, and
\begin{equation}
\label{eq12}
 \hat{\mathbf{E}}({\bf r})
 \!=\!\sum_{\lambda={e},{m}}
 \int_0^{\infty}\!\mathrm{d}\omega\,\int\mathrm{d}^3r'\,
 \bm{G}_\lambda(\mathbf{r},\mathbf{r}',\omega)
 \!\cdot\!\hat{\mathbf{f}}_\lambda(\mathbf{r}',\omega)
 +\mathrm{H.c.},
\end{equation}
with $\bm{G}_\lambda(\mathbf{r},\mathbf{r}',\omega)$ being related
to the classical (retarded) Green tensor
$\bm{G}(\mathbf{r},\mathbf{r}',\omega)$ according to
\begin{align}
\label{eq14}
&\bm{G}_e(\mathbf{r},\mathbf{r}',\omega)
 =\mi\,\frac{\omega^2}{c^2}
 \sqrt{\frac{\hbar}{\pi\varepsilon_0}\,
 \mathrm{Im}\,\varepsilon(\mathbf{r}',\omega)}\,
 \bm{G}(\mathbf{r},\mathbf{r}',\omega),\\[1ex]
\label{eq15}
&\bm{G}_m(\mathbf{r},\mathbf{r}',\omega)
 =\mi\,\frac{\omega}{c}
 \sqrt{-\frac{\hbar}{\pi\varepsilon_0}\,
 \mathrm{Im}\,\kappa(\mathbf{r}',\omega)}
 \bigl[\bm{\nabla}'
 \!\!\times\!\bm{G}(\mathbf{r}',\mathbf{r},\omega)
 \bigr]^{\!\mathsf{T}}.
\end{align}
Note that $\hat{\mathbf{E}}(\hat{\mathbf{r}})$ has the physical
meaning of a displacement field with respect to the atomic
polarization, and the body-assisted induction field reads
\begin{multline}
\label{2.31}
 \hat{\vect{B}}(\vect{r})
 =\sum_{\lambda=e,m}
 \int\dif^3r'\int_0^\infty\frac{\dif\omega}{\mi\omega}\\
 \times\vect{\nabla}\vprod\ten{G}_\lambda(\vect{r},\vect{r}',\omega)
 \sprod\hat{\vect{f}}_\lambda(\vect{r}',\omega)
 +\mathrm{H.c.}
\end{multline}

For magneto-electric bodies, the Green tensor is defined by the
differential equation
\begin{equation}
\label{eq16}
\left[\bm{\nabla}\times \kappa(\mathbf{r},\omega)\bm{\nabla}\times
 \,-\,\frac{\omega^2}{c^2}\,\varepsilon(\mathbf{r},\omega)\right]
 \bm{G}(\mathbf{r},\mathbf{r}',\omega)
 =\bm{\delta}(\mathbf{r}-\mathbf{r}')
\end{equation}
[$\kappa({\bf r},\omega)$ $\!=$ $\!\mu^{-1}({\bf r},\omega)$] together
with the boundary condition
\begin{equation}
\label{eq17}
\bm{G}(\mathbf{r},\mathbf{r}',\omega)\to 0
\quad\mbox{for }|\mathbf{r}-\mathbf{r}'|\to \infty.
\end{equation}
It has the following useful properties \cite{0002}:
\begin{equation}
\label{eq18}
\bm{G}^{\ast}(\mathbf{r},\mathbf{r}',\omega)
 =\bm{G}(\mathbf{r},\mathbf{r}',-\omega^{\ast}),
\end{equation}
\begin{equation}
\label{eq19}
\bm{G}(\mathbf{r},\mathbf{r}',\omega)
 =\bm{G}^\mathsf{T}\!(\mathbf{r}',\mathbf{r},\omega),
\end{equation}
\begin{multline}
\label{eq20}
\int\!\mathrm{d}^3 s\,\Bigl\{\frac{\omega^2}{c^2}\,
 \mathrm{Im}\,\varepsilon(\mathbf{s},\omega)
 \,\bm{G}(\mathbf{r},\mathbf{s},\omega)
 \!\cdot\!\bm{G}^\ast(\mathbf{s},\mathbf{r}',\omega)\\
-\mathrm{Im}\,\kappa(\mathbf{s},\omega)
 \bigl[{\bm{\nabla}}_{\mathbf{s}}\!\times\!
 \bm{G}(\mathbf{s},\mathbf{r},\omega)\bigr]^\mathsf{\!T}
 \!\!\cdot\!
 \bigl[{\bm{\nabla}}_{\mathbf{s}}\!\times\!
 \bm{G}^\ast(\mathbf{s},\mathbf{r}',\omega)\bigr]\Bigr\}\\
=\mathrm{Im}\,\bm{G}(\mathbf{r},\mathbf{r}',\omega).
\end{multline}
Combination of Eq.~(\ref{eq20}) with Eqs.~(\ref{eq14}) and
(\ref{eq15}) leads to
\begin{multline}
\label{eq21}
\sum_{\lambda={e},{m}}\int\mathrm{d}^3 s\,
 \bm{G}_\lambda(\mathbf{r},\mathbf{s},\omega)\!\cdot\!
 \bm{G}^{\ast\mathsf{T}}_\lambda\!(\mathbf{r}',\mathbf{s},\omega)\\
=\frac{\hbar\mu_0}{\pi}\,\omega^2\mathrm{Im}\,
 \bm{G}(\mathbf{r},\mathbf{r}',\omega).
\end{multline}
For an extension of the quantization scheme to arbitrary
linear bodies (including nonlocally responding ones), we refer the
reader to Ref.~\cite{0751}.


\subsection{Atom--field coupling}
\label{sec2.1}

Let us assume that a single atomic transition is coupled
near-resonantly to a body-assisted electromagnetic narrow-band field.
In this case, it is appropriate to use the model of a two-level atom,
where only two atomic energy eigenstates, denoted by $|1\rangle$ and
$|0\rangle$, are involved in the atom--field interaction, so that the
atomic Hamiltonian (\ref{eq2}) effectively reduces to
\begin{equation}
\label{eq22}
\hat{H}_\mathrm{A}
={\textstyle\frac{1}{2}}\hbar\omega_{10}\hat{\sigma}_z
 +{\textstyle\frac{1}{2}}(E_0+E_1)\hat{I}
\end{equation}
[$\omega_{10}$ $\!=$ $\!(E_1$ $\!-$ $\!E_0)/\hbar$; $\hat{\sigma}_z$
$\!=$ $\!|1\rangle\langle 1|$ $\!-$ $\!|0\rangle\langle 0|$;
$\hat{I}$: unit operator], and the electric dipole moment~(\ref{eq10})
takes the form
\begin{equation}
\label{eq23}
\hat{\mathbf{d}}=\mathbf{d}_{01}\hat{\sigma}+\mathrm{H.c.}
\end{equation}
($\mathbf{d}_{mn}$ $\!=$ $\!\langle m|\hat{\mathbf{d}}|n\rangle$,
$\mathbf{d}_{mm}$ $\!=$ $\!\mathbf{0}$, $\hat{\sigma}$ $\!=$
$\!|0\rangle\langle1|$). Consequently, upon recalling
Eq.~(\ref{eq12}), the interaction Hamiltonian (\ref{eq9}) can be
written as
\begin{multline}
\label{eq24}
\hspace*{-2ex}
\hat{H}_\mathrm{AF}=\\
-\!\!\sum_{\lambda=\mathrm{e},\mathrm{m}}
 \!\int_0^{\infty}\!\!\mathrm{d}\omega
 \!\!\int\!\mathrm{d}^3r\,
 \bigl(\mathbf{d}_{10}\hat{\sigma}^\dagger
 +\mathbf{d}_{01}\hat{\sigma}\bigr)
 \!\cdot\!\bm{G}_\lambda(\mathbf{r}_\mathrm{A},\mathbf{r},\omega)
 \!\cdot\!\hat{\mathbf{f}}_\lambda(\mathbf{r},\omega)\\
+\mathrm{H.c.}
\end{multline}

For the following, it is useful to introduce position-dependent
photon-like creation and annihilation operators
$\hat{a}(\mathbf{r},\omega)$ and $\hat{a}^\dagger(\mathbf{r},\omega)$
according to the definition
\begin{multline}
\label{eq25}
\hspace*{-2ex}
\hat{a}(\mathbf{r},\omega)=\\
\hspace{.5ex}
 -\frac{1}{\hbar g(\mathbf{r},\omega)}
 \sum_{\lambda=e,m}\int\mathrm{d}^3r'\,
 \mathbf{d}_{10}\!\cdot\!\bm{G}_\lambda(\mathbf{r},\mathbf{r}',\omega)
 \!\cdot\!\hat{\mathbf{f}}_\lambda(\mathbf{r}',\omega),
\end{multline}
where
\begin{equation}
\label{eq27}
g(\mathbf{r},\omega)=
\sqrt{\frac{\mu_0}{\hbar\pi}\,\omega^2
 \mathbf{d}_{10}\!\cdot\!
 \mathrm{Im}\,\bm{G}(\mathbf{r},\mathbf{r},\omega)
 \!\cdot\!\mathbf{d}_{01}}\,.
\end{equation}
Substituting Eq.~(\ref{eq25}) into Eq.~(\ref{eq24}) and restricting
our attention to the rotating-wave approximation, we may write the
interaction Hamiltonian in the form
\begin{equation}
\label{eq34}
\hat{H}_\mathrm{AF}
=\hbar\int_0^{\infty}\!\mathrm{d}\omega\,
 g(\mathbf{r}_\mathrm{A},\omega)
 \hat{a}(\mathbf{r}_\mathrm{A},\omega)
 \hat{\sigma}^\dagger+\mathrm{H.c.},
\end{equation}
showing that $g(\mathbf{r},\omega)$ may be regarded as a generalized
atom--field coupling strength.

The commutation relations of $\hat{a}(\mathbf{r},\omega)$ and
$\hat{a}^\dagger(\mathbf{r},\omega)$ can be found from the
commutation relations (\ref{eq7}) and (\ref{eq8}) by making use of
the definitions (\ref{eq25}) and (\ref{eq27}) and the integral
relation~(\ref{eq21}), resulting in
\begin{align}
\label{eq28}
&\left[\hat{a}(\mathbf{r},\omega),
 \hat{a}^\dagger(\mathbf{r}',\omega')\right]
=\frac{g(\mathbf{r},\mathbf{r}',\omega)}
 {g(\mathbf{r},\omega)g(\mathbf{r}',\omega)}\,\delta(\omega-\omega'),
 \\[1ex]
\label{eq29}
&\left[\hat{a}(\mathbf{r},\omega),\hat{a}(\mathbf{r},\omega')\right]
 =0,
\end{align}
with
\begin{equation}
\label{eq28.1}
g(\mathbf{r},\mathbf{r}',\omega)
=\frac{\mu_0}{\hbar\pi}\,\omega^2\mathbf{d}_{10}
 \!\cdot\!\mathrm{Im}\,\bm{G}(\mathbf{r},\mathbf{r}',\omega)
 \!\cdot\!\mathbf{d}_{01}.
\end{equation}
Note that
\begin{equation}
\label{eq28.2}
\left[\hat{a}(\mathbf{r},\omega),
 \hat{a}^\dagger(\mathbf{r},\omega')\right]
 =\delta(\omega-\omega').
\end{equation}

The definition of the ground state $|\{0\}\rangle$ of the system
composed of the bodies and the electromagnetic field,
$\hat{\mathbf{f}}_\lambda(\mathbf{r},\omega)|\{0\}\rangle$ $\!=$
$\!0$ ($\forall$ $\!\lambda,\mathbf{r},\omega$), implies that
\begin{equation}
\label{eq30b}
\hat{a}(\mathbf{r},\omega)|\{0\}\rangle
=0\quad\forall\ \mathbf{r},\omega.
\end{equation}
The operators $\hat{a}^\dagger(\mathbf{r},\omega)$ can be used to
define single-quantum excitation states
\begin{equation}
\label{eq30}
|\mathbf{r},\omega\rangle
 =\hat{a}^\dagger(\mathbf{r},\omega)|\{0\}\rangle.
\end{equation}
{F}rom Eq.~(\ref{eq28}) it then follows that these states are
orthogonal with respect to frequency, but not with respect to
position,
\begin{equation}
\label{eq32.1}
\langle\mathbf{r},\omega|\mathbf{r}',\omega'\rangle
=\frac{g(\mathbf{r},\mathbf{r}',\omega)}
 {g(\mathbf{r},\omega)g(\mathbf{r}',\omega)}\,
 \delta(\omega-\omega');
\end{equation}
for equal positions we thus have
\begin{equation}
\label{eq32.2}
\langle\mathbf{r},\omega|\mathbf{r},\omega'\rangle
=\delta(\omega-\omega')
\end{equation}
[cf.~Eq.~(\ref{eq28.2})]. Furthermore, the states $|\mathbf{r},\omega\rangle$
are eigenstates of $\hat{H}_\mathrm{F}$ carrying one quantum of energy
$\hbar\omega$,
\begin{equation}
\label{eq32}
\hat{H}_\mathrm{F}|\mathbf{r},\omega\rangle
 =\hbar\omega|\mathbf{r},\omega\rangle ,
\end{equation}
as can be seen from the commutation relation
\begin{equation}
\label{eq31}
\bigl[\hat{H}_\mathrm{F},\hat{a}^\dagger(\mathbf{r},\omega)\bigr]
 =\hbar\omega\hat{a}^\dagger(\mathbf{r},\omega),
\end{equation}
which is a consequence of Eqs.~(\ref{eq6}) and (\ref{eq25})
together with the commutation relations (\ref{eq7}) and (\ref{eq8}).

Equation~(\ref{eq34}) implies that the states
$|\mathbf{r}_\mathrm{A},\omega\rangle$ \mbox{($\omega$ $\!\simeq$
$\!\omega_{10}$)} as given according to Eq.~(\ref{eq30}) describe the
single-quantum excited states of the body-assisted electromagnetic
field which may be obtained after an atom positioned at
$\mathbf{r}_\mathrm{A}$ has undergone an (electric-dipole) transition
$|1\rangle$ $\!\rightarrow$ $\!|0\rangle$. Hence, these states can be
regarded as spanning the state space of all the states that are
resonantly coupled to an initially excited two-level atom via
$\hat{H}_\mathrm{AF}$ as given by Eq.~(\ref{eq34}), which is why it
will be sufficient to work with them in the following.


\section{The Casimir--Polder force}
\label{sec3}

As shown by Casimir and Polder \cite{0030}, dispersion forces on
ground-state systems can be obtained by means of time-independent
perturbation theory, by calculating the shift of the respective
(unperturbed) ground-state energy due to the atom--field coupling for
given atomic center-of-mass position,
\begin{equation}
\label{eq21.1}
\Delta E=\Delta E^{(0)}+U(\mathbf{r}_\mathrm{A}),
\end{equation}
and identifying its position-dependent part
$U(\mathbf{r}_\mathrm{A})$
as the potential from which the force can be derived,
\begin{equation}
\label{eq21.2}
\mathbf{F}(\mathbf{r}_\mathrm{A})
=-\bm{\nabla}_{\!\!\mathrm{A}}U(\mathbf{r}_\mathrm{A}).
\end{equation}
When dealing with a system prepared in an excited state, this simple
approach may fail for two reasons. Firstly, excited states undergo a
temporal evolution, so that a dynamic approach to the CP force may be
more appropriate than a static one. Secondly, if an excited atom
strongly interacts with a body-assisted electromagnetic narrow-band
field, perturbation theory may break down.

In the following, we focus on the CP force that results from resonant
single-quantum exchange between an atom and a body-assisted field,
which may be described by approximating the atom by a two-level system
and treating the atom--field interaction in rotating-wave
approximation. We will first (Sec.~\ref{sec3.1}) develop a simplified
approximate theory that is similar in spirit with both Casimir and
Polder's original work \cite{0030} and the quasi-stationary
dressed-state approach to strong atom--field coupling
\cite{0179,0407}. We will then (Sec.~\ref{sec3.2}) develop a more
complete dynamic theory, by starting from the operator-valued Lorentz
force \cite{0008,0012} and evaluating its time-dependent expectation
value by solving the Schr\"{o}dinger equation for chosen initial
preparation, thereby taking into account the finite bandwidth of a
realistic body-assisted electromagnetic field.


\subsection{Static approximation}
\label{sec3.1}

As already mentioned at the end of Sec.~\ref{sec2.1}, the Hilbert
space for the case where a two-level atom that is initially prepared
in the upper state resonantly interacts with a body-assisted
electromagnetic field can be spanned by the states
$|1\rangle|\{0\}\rangle$ and
$|0\rangle|\mathbf{r}_\mathrm{A},\omega\rangle$. Using Eq.~(\ref{eq1})
[together with Eqs.~(\ref{eq6}), (\ref{eq22}), and (\ref{eq34})] and
recalling the definition (\ref{eq30}) and the commutation relations
(\ref{eq28}) and (\ref{eq32}), we derive
\begin{multline}
\label{eq34.1}
\hat{H}|1\rangle|\{0\}\rangle
 = E_1|1\rangle|\{0\}\rangle\\[.5ex]
+\hbar\int_0^\infty\mathrm{d}\omega\,g(\mathbf{r}_\mathrm{A},\omega)
 |0\rangle|\mathbf{r}_\mathrm{A},\omega\rangle,
\end{multline}
\begin{multline}
\label{eq34.2}
\hat{H}|0\rangle|\mathbf{r}_\mathrm{A},\omega\rangle
=(E_0+\hbar\omega)
 |0\rangle|\mathbf{r}_\mathrm{A},\omega\rangle\\[.5ex]
+\hbar g(\mathbf{r}_\mathrm{A},\omega)|1\rangle|\{0\}\rangle.
\end{multline}
Let us assume that the body-assisted field that resonantly interacts
with the atom can be approximated by a single Lorentzian-type
nonmonochromatic mode $\nu$ of mid-frequency $\omega_\nu$ and full
width at half maximum (FWHM)~$\gamma_\nu$,
\begin{equation}
\label{eq35}
g^2(\mathbf{r},\omega)
 =g^2(\mathbf{r},\omega_\nu)\,
 \frac{\gamma_\nu^2/4}{(\omega-\omega_\nu)^2+\gamma_\nu^2/4}\,,
\end{equation}
such that $\gamma_\nu\to 0$ and $g^2(\mathbf{r},\omega_\nu)\to\infty$,
but the product $\gamma_\nu g^2(\mathbf{r},\omega_\nu)$ remains
finite, thus
\begin{equation}
\label{eq35-1}
g^2(\mathbf{r},\omega)
\to \textstyle{\frac{1}{2}}\pi\gamma_\nu
g^2(\mathbf{r},\omega_\nu)\delta(\omega-\omega_\nu)
\end{equation}
in the limiting case of a monochromatic mode.

\begin{figure}[!t!]
\noindent
\begin{center}
\includegraphics[width=0.6\linewidth]{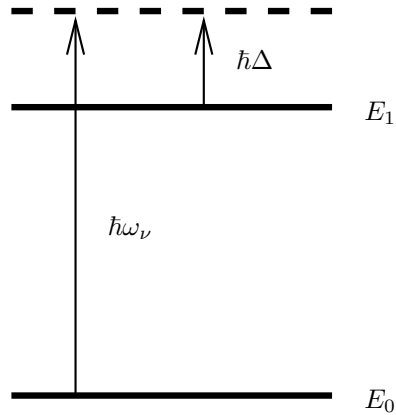}
\end{center}
\caption{
\label{fig1}
Level scheme of a two-level atom interacting with 
a single-resonance electromagnetic field.
}
\end{figure}

The single-resonance assumption implies that Eq.~(\ref{eq34.1}) takes
the form
\begin{equation}
\label{eq36}
\hat{H}|1\rangle|\{0\}\rangle=E_1|1\rangle|\{0\}\rangle
 +{\textstyle\frac{1}{2}}
 \hbar\Omega_\mathrm{R}(\mathbf{r}_\mathrm{A})|0\rangle|1_\nu\rangle,
\end{equation}
where
\begin{equation}
\label{eq39.1}
\Omega_\mathrm{R}(\mathbf{r}_\mathrm{A})
=\sqrt{2\pi\gamma_\nu g^2(\mathbf{r}_\mathrm{A},\omega_\nu)}
\end{equation}
is the vacuum Rabi frequency, and the state $|1_\nu\rangle$
is defined by
\begin{equation}
\label{eq36.1}
|1_\nu\rangle
 =\sqrt{\frac{\gamma_\nu}{2\pi}}
 \int_{\omega_\nu-\delta\omega/2}^{\omega_\nu+\delta\omega/2}
\frac{\mathrm{d}\omega}
 {\sqrt{(\omega-\omega_\nu)^2+\gamma_\nu^2/4}}\,
 |\mathbf{r}_\mathrm{A},\omega\rangle,
\end{equation}
where $\delta\omega$ is a measure of the distance between two
neighboring lines ($\gamma_\nu$ $\!\ll$ $\!\delta\omega$). On
recalling Eq.~(\ref{eq32.2}), it is not difficult to see that
(for $\gamma_\nu\to 0$) it is normalized to unity,
\begin{equation}
\label{eq36.2}
\langle 1_\nu|1_\nu\rangle=1.
\end{equation}
Within the approximation scheme used, from Eq.~(\ref{eq34.2}) it
follows that
\begin{equation}
\label{eq37}
\hat{H}|0\rangle|1_\nu\rangle
 =(E_0\!+\!\hbar\omega_\nu)|0\rangle|1_\nu\rangle
 +{\textstyle\frac{1}{2}}\hbar
 \Omega_\mathrm{R}(\mathbf{r}_\mathrm{A})
 |1\rangle|\{0\}\rangle.
\end{equation}
Hence we are left with Eqs.~(\ref{eq36}) and (\ref{eq37}) revealing
that the subspace spanned by the states $|1\rangle|\{0\} \rangle$ and
$|0\rangle|1_\nu\rangle$ is invariant under $\hat{H}$ which, on this
subspace, takes the well-known Jaynes--Cummings form
\begin{equation}
\label{eq38}
H_{\alpha\beta}=
 \left\lgroup\begin{array}{cc}
 E_1
 &{\textstyle\frac{1}{2}}\hbar
 \Omega_\mathrm{R}(\mathbf{r}_\mathrm{A})
 \\[.5ex]
 {\textstyle\frac{1}{2}}\hbar\Omega_\mathrm{R}(\mathbf{r}_\mathrm{A})
 &E_0+\hbar\omega_\nu
 \end{array}\right\rgroup
\end{equation}
(for the level scheme, see Fig.~\ref{fig1}).

Obviously, the single-resonance assumption together with the
approximation $\gamma_\nu\to 0$ implies that the coupling of the atom
to the residual continuum of the body-assisted field is fully ignored,
i.e., both radiative and nonradiative losses are disregarded.
This means that the results found in this way are only valid on time
scales which are short with respect to the associated relaxation
times, as will explicitly be shown in Sec.~\ref{sec3.2}.

Straightforward diagonalization of the Jaynes--Cum\-mings
Hamiltonian~(\ref{eq38}) yields the two eigenenergies
\begin{equation}
\label{eq40}
E_\pm={\textstyle\frac{1}{2}}(E_0+E_1+\hbar\omega_\nu)
 \pm{\textstyle\frac{1}{2}}\hbar\Omega(\mathbf{r}_\mathrm{A})
\end{equation}
with
\begin{equation}
\label{eq43c}
\Omega(\mathbf{r}_\mathrm{A})
=\sqrt{\Omega_\mathrm{R}^2(\mathbf{r}_\mathrm{A})+\Delta^2}\,,
\end{equation}
where
\begin{equation}
\label{eq39}
\Delta=\omega_\nu-\omega_{10}
\end{equation}
is the atom--field detuning, and the eigenstates (the dressed states)
can be written in the form
\begin{align}
\label{eq42}
&|+\rangle=\cos[\theta_c(\mathbf{r}_\mathrm{A})]\,
|1\rangle|\{0\}\rangle
 +\sin[\theta_c(\mathbf{r}_\mathrm{A})]\,
 |0\rangle|1_\nu\rangle,
\\[1ex]
\label{eq42.1}
&|-\rangle=-\sin[\theta_c(\mathbf{r}_\mathrm{A})]\,
|1\rangle|\{0\}\rangle
 +\cos[\theta_c(\mathbf{r}_\mathrm{A})]\,
 |0\rangle|1_\nu\rangle,
\end{align}
where, according to
\begin{equation}
\label{eq41}
\tan[2\theta_c(\mathbf{r}_\mathrm{A})]
=-\frac{\Omega_\mathrm{R}(\mathbf{r}_\mathrm{A})}{\Delta}\,,
 \quad\theta_c(\mathbf{r}_\mathrm{A})\in[0,\pi/2],
\end{equation}
the coupling angle $\theta_c(\mathbf{r}_\mathrm{A})$ has been
introduced, and the relation
\begin{equation}
\label{eq41.1}
\frac{1}{\sqrt{1+\cot^2(\alpha)}}=\sin(\alpha),
 \quad\alpha\in[0,\pi]
\end{equation}
has been used.
Note that in the case of exact resonance, $\Delta$ $\!=$ $\!0$,
Eq.~(\ref{eq43c}) reduces to
\begin{equation}
\label{eq43c-1}
\Omega(\mathbf{r}_\mathrm{A})
= \Omega_\mathrm{R}(\mathbf{r}_\mathrm{A}),
\end{equation}
and Eqs.~(\ref{eq42}) and (\ref{eq42.1}) simplify to
\begin{equation}
\label{eq42b}
|\pm\rangle=\frac{1}{\sqrt{2}}\,(\pm|1\rangle|\{0\}\rangle
 +|0\rangle|1_\nu\rangle).
\end{equation}

Comparing Eq.~(\ref{eq40}) with Eq.~(\ref{eq21.1}), we can conclude
that for a system prepared in the state $|+\rangle$ or $|-\rangle$,
respectively, the CP potential $U_+(\mathbf{r}_\mathrm{A})$ or
$U_-(\mathbf{r}_\mathrm{A})$ is given by
\begin{equation}
\label{eq43}
U_\pm(\mathbf{r}_\mathrm{A})
=\pm{\textstyle\frac{1}{2}}\hbar\Omega(\mathbf{r}_\mathrm{A}),
\end{equation}
from which the associated CP force
\begin{equation}
\label{eq43b}
\mathbf{F}_\pm(\mathbf{r}_\mathrm{A})
=-\bm{\nabla}_{\!\!\mathrm{A}}U_\pm(\mathbf{r}_\mathrm{A})
\end{equation}
can be obtained. Equation (\ref{eq43}) generalizes the result derived
in Ref.~\cite{0179} for the case of a two-level atom strongly coupled
to a single standing wave in a cavity to an arbitrary resonator-like
arrangement of bodies giving rise to strong atom--field coupling.
As can be seen from Eq.~(\ref{eq43c}) together with
Eqs.~(\ref{eq39.1}), and (\ref{eq28.1}), the CP potential can be quite
generally expressed in terms of the imaginary part of
the Green tensor of the bodies as
\begin{multline}
\label{eq43b-1}
U_\pm(\mathbf{r}_\mathrm{A}) =
\\
\pm {\textstyle\frac{1}{2}}
\sqrt{2\hbar\mu_0\gamma_\nu\omega_\nu^2
\mathbf{d}_{10}\!\cdot\!
\mathrm{Im}\,\bm{G}(\mathbf{r}_\mathrm{A},\mathbf{r}_\mathrm{A},
\omega_\nu)\!\cdot\!\mathbf{d}_{10}
+\hbar^2\Delta^2}\,.
\end{multline}

At this point we recall that the ground-state variance of the electric
field in a frequency interval $\delta\omega$ is given by~\cite{0012}
\begin{align}
\label{eq43.1}
\bigl\langle[\Delta\hat{\mathbf{E}}(\mathbf{r})]^2
\bigr\rangle_{\delta\omega}
&=\Bigl[\langle\{0\}|\hat{\mathbf{E}}^2(\mathbf{r})|\{0\}\rangle-
\langle\{0\}|\hat{\mathbf{E}}(\mathbf{r})|\{0\}\rangle^2
\Bigr]_{\delta\omega}
\nonumber\\[.5ex]
&=\frac{\hbar\mu_0}{\pi}\int_{\delta\omega}
\mathrm{d}\omega\,
\omega^2\,\mathrm{Im}\,[\mathrm{Tr}\,
\bm{G}(\mathbf{r},\mathbf{r},\omega)],
\end{align}
which in the single-mode case considered here reduces to
\begin{equation}
\label{eq43.1x}
\langle[\Delta\hat{\mathbf{E}}(\mathbf{r})]^2\rangle
_{\delta\omega}
= {\textstyle\frac{1}{2}}\hbar\mu_0\gamma_\nu
\omega_\nu^2\,\mathrm{Im}\,[\mathrm{Tr}\,
\bm{G}(\mathbf{r},\mathbf{r},\omega_\nu)].
\end{equation}
Comparing Eq.~(\ref{eq43.1x}) with Eq.~(\ref{eq43}) [recall
Eqs.~(\ref{eq27}) and (\ref{eq39.1})], we see that the CP potential is
essentially determined by the ground-state fluctuations of the
electric field at the position of the atom along the direction of its
transition dipole moment. Moreover, we see that for a system prepared
in the state $|+\rangle$ ($|-\rangle$), the atom is repelled from
(attracted towards) regions of high field fluctuations. This is in
agreement with the result found for the case of a standing wave in a
cavity \cite{0179}, where the atom is repelled from (attracted
towards) the antinodes when the system is in the state $|+\rangle$
($|-\rangle$).

To make contact with the results of perturbation theory, which are
applicable in the case of weak atom--field coupling, let us consider
the limiting case of large detuning, $|\Delta|$ $\!\gg$
$\!\Omega_\mathrm{R}(\mathbf{r}_\mathrm{A})$, where, according to
Eq.~(\ref{eq41}), the coupling angle approaches
$\theta_c(\mathbf{r}_\mathrm{A})$ $\!=$ $\!\pi/2$ and
$\theta_c(\mathbf{r}_\mathrm{A})$ $\!=$ $\!0$, respectively, for
positive and negative detuning, and the dressed states~(\ref{eq42})
and (\ref{eq42.1}) approximate to
\begin{align}
\label{eq42c}
&|+\rangle=\begin{cases}\hspace{1.75ex}|0\rangle|1_\nu\rangle
 &\mbox{for }\Delta>0,\\
 -|1\rangle|\{0\}\rangle
 &\mbox{for }\Delta<0,\end{cases}\\
\label{eq42.1c}
&|-\rangle=\begin{cases}\hspace{1.75ex}|1\rangle|\{0\}\rangle
 &\mbox{for }\Delta>0,\\
 \hspace{1.75ex}|0\rangle|1_\nu\rangle
 &\mbox{for }\Delta<0.\end{cases}
\end{align}
By expanding the square root in Eq.~(\ref{eq43c}) and recalling
Eq.~(\ref{eq21.1}), one finds that the CP potentials
$U_{1\{0\}}(\mathbf{r}_\mathrm{A})$ and
$U_{01_\nu}(\mathbf{r}_\mathrm{A})$ associated with
the states $|1\rangle|\{0\}\rangle$ and $|0\rangle|1_\nu\rangle$,
respectively, can be written as
\begin{align}
\label{eq43.1g2}
&U_{1\{0\}}(\mathbf{r}_\mathrm{A})
 =-\frac{\hbar}{4\Delta}\,
 \Omega^2_\mathrm{R}(\mathbf{r}_\mathrm{A}),\\
\label{eq43.1h2}
&U_{01_\nu}(\mathbf{r}_\mathrm{A})
 =\frac{\hbar}{4\Delta}\,
 \Omega^2_\mathrm{R}(\mathbf{r}_\mathrm{A}).
\end{align}
Next, we employ the Kramers--Kronig relations for the (spectral)
response function $\omega^2\bm{G}^{(1)}(\mathbf{r},\mathbf{r},\omega)$
to note that, according to the single-resonance
approximation~(\ref{eq35-1}) made, we may write [recall
Eqs.~(\ref{eq27}) and (\ref{eq39.1})]
\begin{multline}
\label{eq43.1i}
\mu_0\omega^2
 \mathbf{d}_{10}\!\cdot\!\mathrm{Re}\,
 \bm{G}^{(1)}(\mathbf{r},\mathbf{r},\omega)
 \!\cdot\!\mathbf{d}_{01}\\
=\frac{\mu_0}{\pi}\,
 \mathcal{P}\int_{-\infty}^\infty
 \frac{\mathrm{d}\omega'}{\omega'-\omega}\,
 \omega^{\prime 2}\mathbf{d}_{10}\!\cdot\!\mathrm{Im}\,
 \bm{G}^{(1)}(\mathbf{r},\mathbf{r},\omega')
 \!\cdot\!\mathbf{d}_{01}\\
\to {\textstyle\frac{1}{2}}\pi\hbar\gamma_\nu
 g^2(\mathbf{r}_\mathrm{A},\omega_\nu)
 \mathcal{P}\int_{-\infty}^\infty
 \frac{\mathrm{d}\omega'}{\omega'-\omega}\,
 \delta(\omega-\omega_\nu)\\
=\frac{\hbar\Omega^2_\mathrm{R}(\mathbf{r})}{4(\omega_\nu-\omega)}
 \quad\mbox{for }\omega\neq\omega_\nu
\end{multline}
($\mathcal{P}$: principal value), which is valid up to a
position-independent constant. Here, the Green tensor has been
decomposed into the translationally invariant bulk part $\bm{G}^{(0)}$
and the scattering part $\bm{G}^{(1)}$,
\begin{equation}
\label{eq64}
\bm{G}(\mathbf{r},\mathbf{r}',\omega)
 =\bm{G}^{(0)}(\mathbf{r},\mathbf{r}',\omega)
 +\bm{G}^{(1)}(\mathbf{r},\mathbf{r}',\omega).
\end{equation}
Upon discarding position-independent terms, Eqs.~(\ref{eq43.1g2}) and
(\ref{eq43.1h2}) can hence be rewritten in the form
\begin{align}
\label{eq43.1g}
&U_{1\{0\}}(\mathbf{r}_\mathrm{A})
 =-\mu_0\omega_{10}^2
 \mathbf{d}_{10}\!\cdot\!\mathrm{Re}\,
 \bm{G}^{(1)}(\mathbf{r}_\mathrm{A},
 \mathbf{r}_\mathrm{A},\omega_{10})
 \!\cdot\!\mathbf{d}_{01},\\[.5ex]
\label{eq43.1h}
&U_{01_\nu}(\mathbf{r}_\mathrm{A})
 =\mu_0\omega_{10}^2
 \mathbf{d}_{10}\!\cdot\!\mathrm{Re}\,
 \bm{G}^{(1)}(\mathbf{r}_\mathrm{A},
 \mathbf{r}_\mathrm{A},\omega_{10})
 \!\cdot\!\mathbf{d}_{01}.
\end{align}

As expected, $U_{1\{0\}}(\mathbf{r}_\mathrm{A})$, Eq.~(\ref{eq43.1g}),
is just the resonant part of the CP potential obtained in
leading-order perturbation theory for a two-level atom that is
prepared in the upper state and (weakly) coupled to a body-assisted
electromagnetic ground state, cf.~Refs.~\cite{0012,0008}. The fact
that only the resonant part of the potential appears is obviously due
to the rotating-wave approximation. The result (\ref{eq43.1h}) is new,
$U_{01_\nu}(\mathbf{r}_\mathrm{A})$ gives the resonant part of the
(weak-coupling) CP potential for the case where instead of the atom
being excited and the field being in its ground state, the atom is in
its ground state, but the field is in single-photon Fock
state~(\ref{eq36.1}). It is seen that the potentials
$U_{01_\nu}(\mathbf{r}_\mathrm{A})$ and
$U_{1\{0\}}(\mathbf{r}_\mathrm{A})$ and hence also the associated
forces just differ in sign. Regarding the forces as being the result
of recoil due to emission/absorption of a photon (where in the
presence of bodies the photon is predominantly emitted/absorbed in a
certain direction, leading to a nonvanishing net recoil), this
difference obviously reflects the fact that for the excited atom the
resonant process responsible for the potential is the emission of a
photon while for the field being excited the resonant process is the
absorption of a photon (hence the direction of net recoil is
reversed with respect to the emission case). It should be pointed out
that the validity of Eq.~(\ref{eq43.1g}) does not require a
single-mode approximation, so this equation is generally valid in the
weak-coupling regime. Clearly, the state $|1\rangle|\{0\}\rangle$ can
then no longer be regarded as being a (quasi-)stationary one in
general.

Finally, let us consider the situation where the system is prepared in
a superposition state
\begin{equation}
\label{eq43.2}
|\theta\rangle\!=\!\cos\theta\,|1\rangle|\{0\}\rangle
 \!+\!\sin\theta\,|0\rangle|1_\nu\rangle,
\end{equation}
$\theta\in[0,\pi]$, which includes the special cases
\begin{align}
\label{eq43.3}
&|\theta\!=\!0\rangle=|1\rangle|\{0\}\rangle,\\
\label{eq43.3b}
&|\theta\!=\!\pi/2\rangle=
|0\rangle|1_\nu\rangle,\\
\label{eq43.3c}
&|\theta\!=\!\theta_c\rangle=|+\rangle,\\
\label{eq43.3d}
&|\theta\!=\!\theta_c\!+\!\pi/2\rangle=|-\rangle.
\end{align}
Note that when the system is prepared in a state $|\theta\rangle$
whose projections onto both $|+\rangle$ and $|-\rangle$ do not vanish,
Rabi oscillations will be observed, leading to a nontrivial dynamics
of the CP potential and the associated force, as we will show in
Sec.~\ref{sec3.2}. Here, we restrict our attention to the potential
and the force at initial time, where the system is in the state
$|\theta\rangle$.

The CP potential as the position-dependent part of the expectation
value of the Jaynes--Cummings Hamiltonian of the system prepared in a
state $|\theta\rangle$ is calculated to be
\begin{multline}
\label{eq43.4}
U_\theta(\mathbf{r}_\mathrm{A})
=|\langle\theta|+\rangle|^2U_+(\mathbf{r}_\mathrm{A})
 +|\langle\theta|-\rangle|^2U_-(\mathbf{r}_\mathrm{A})\\
=\cos^2[\theta\!-\!\theta_c(\mathbf{r}_\mathrm{A})]
 U_+(\mathbf{r}_\mathrm{A})
 +\sin^2[\theta\!-\!\theta_c(\mathbf{r}_\mathrm{A})]
 U_-(\mathbf{r}_\mathrm{A})\\
={\textstyle\frac{1}{2}}\hbar
 \cos\bigl\{2[\theta\!-\!\theta_c(\mathbf{r}_\mathrm{A})]\bigr\}
 \Omega(\mathbf{r}_\mathrm{A}),
\end{multline}
where Eqs.~(\ref{eq42}), (\ref{eq42.1}) and (\ref{eq43}) have been
used. According to Eq.~(\ref{eq21.2}), the associated CP force
then follows as
\begin{align}
\label{eq43.5}
\mathbf{F}_\theta(\mathbf{r}_\mathrm{A})
=&\,-{\textstyle\frac{1}{2}}\hbar
 \cos\bigl\{2[\theta\!-\!\theta_c(\mathbf{r}_\mathrm{A})]\bigr\}
 \bm{\nabla}_{\!\!\mathrm{A}}
 \Omega(\mathbf{r}_\mathrm{A})\nonumber\\
&\,-\hbar
 \sin\bigl\{2[\theta\!-\!\theta_c(\mathbf{r}_\mathrm{A})]\bigr\}
 \Omega(\mathbf{r}_\mathrm{A})
 \bm{\nabla}_{\!\!\mathrm{A}}\theta_c(\mathbf{r}_\mathrm{A}).
\end{align}
In order to carry out the derivatives in Eq.~(\ref{eq43.5}), we note
that Eqs.~(\ref{eq43c}), (\ref{eq41}) and (\ref{eq41.1}) imply
\begin{gather}
\label{eq43.6}
\sin[2\theta_c(\mathbf{r}_\mathrm{A})]\Omega(\mathbf{r}_\mathrm{A})
 =\Omega_\mathrm{R}(\mathbf{r}_\mathrm{A}),\\[1ex]
\label{eq43.7}
\bm{\nabla}_{\!\!\mathrm{A}}\Omega(\mathbf{r}_\mathrm{A})
 =\sin[2\theta_c(\mathbf{r}_\mathrm{A})]
 \bm{\nabla}_{\!\!\mathrm{A}}\Omega_\mathrm{R}
 (\mathbf{r}_\mathrm{A}),
\end{gather}
while taking the derivative of Eq.~(\ref{eq41}) and using the
relations
\begin{equation}
\label{eq73.16}
\frac{1}{\sqrt{1+\tan^2(\alpha)}}
 =\begin{cases}\cos(\alpha)&
  \mbox{for }\alpha\in[0,\pi/2],\\
   -\cos(\alpha)&
  \mbox{for }\alpha\in[\pi/2,\pi],
\end{cases}
\end{equation}
we derive
\begin{equation}
\label{eq43.7b}
\bm{\nabla}_{\!\!\mathrm{A}}\theta_c(\mathbf{r}_\mathrm{A})
=-\frac{\cos^2[2\theta_c(\mathbf{r}_\mathrm{A})]
 \bm{\nabla}_{\!\!\mathrm{A}}\Omega_\mathrm{R}
 (\mathbf{r}_\mathrm{A})}
 {2\Delta}\,.
\end{equation}
Using Eqs.~(\ref{eq41}), (\ref{eq43.6}), (\ref{eq43.7}), and
(\ref{eq43.7b}), we can eventually write Eq.~(\ref{eq43.5}) in the
form
\begin{multline}
\label{eq43.8}
\mathbf{F}_\theta(\mathbf{r}_\mathrm{A})
=-{\textstyle\frac{1}{2}}\hbar\left(
 \cos\bigl\{2[\theta\!-\!\theta_c(\mathbf{r}_\mathrm{A})]\bigr\}
 +\cot[2\theta_c(\mathbf{r}_\mathrm{A})]\right.\\
\left.\times
 \sin\bigl\{2[\theta\!-\!\theta_c(\mathbf{r}_\mathrm{A})]\bigr\}
 \right)
 \bm{\nabla}_{\!\!\mathrm{A}}\Omega(\mathbf{r}_\mathrm{A}),
\end{multline}
which of course contains the previous result (\ref{eq43b}) in the
special case of the system being prepared in one of the dressed
states $|\pm\rangle$, recall Eqs.~(\ref{eq43.3c}) and
(\ref{eq43.3d}).

In the case of exact resonance, $\Delta$ $\!=$ $\!0$, where
\mbox{$\theta_\mathrm{c}$ $\!=$ $\!\pi/4$}, Eqs.~(\ref{eq43.4}) and
(\ref{eq43.8}) reduce to
\begin{equation}
\label{eq43.9}
U_\theta(\mathbf{r}_\mathrm{A})
={\textstyle\frac{1}{2}}\hbar\sin(2\theta)
 \Omega_\mathrm{R}(\mathbf{r}_\mathrm{A})
\end{equation}
and
\begin{equation}
\label{eq43.10}
\mathbf{F}_\theta(\mathbf{r}_\mathrm{A})
=-{\textstyle\frac{1}{2}}\hbar\sin(2\theta)
 \bm{\nabla}_{\!\!\mathrm{A}}
 \Omega_\mathrm{R}(\mathbf{r}_\mathrm{A}),
\end{equation}
respectively. In complete analogy to the derivation of
Eqs.~(\ref{eq43.1g}) and (\ref{eq43.1h}), one can show that for large
detuning, $|\Delta|$ $\!\gg$
$\!\Omega_\mathrm{R}(\mathbf{r}_\mathrm{A})$, which corresponds to
weak atom--field coupling, Eqs.~(\ref{eq43.4}) and (\ref{eq43.8})
approximate to
\begin{multline}
\label{eq43.11}
U_\theta(\mathbf{r}_\mathrm{A})
=-\frac{\hbar}{4\Delta}\,\cos(2\theta)
 \Omega^2_\mathrm{R}(\mathbf{r}_\mathrm{A})\\
= -\cos(2\theta)\mu_0\omega_{10}^2
 \mathbf{d}_{10}\!\cdot\!\mathrm{Re}\,
 \bm{G}^{(1)}(\mathbf{r}_\mathrm{A},
 \mathbf{r}_\mathrm{A},\omega_{10})
 \!\cdot\!\mathbf{d}_{01}
\end{multline}
and
\begin{multline}
\label{eq43.12}
\mathbf{F}_\theta(\mathbf{r}_\mathrm{A})
=\frac{\hbar}{2\Delta}\,\cos(2\theta)
 \Omega_\mathrm{R}(\mathbf{r}_\mathrm{A})\bm{\nabla}_{\!\!\mathrm{A}}
 \Omega_\mathrm{R}(\mathbf{r}_\mathrm{A})\\
=\cos(2\theta)\mu_0\omega_{10}^2\bm{\nabla}_{\!\!\mathrm{A}}
 \mathbf{d}_{10}\!\cdot\!\mathrm{Re}\,
 \bm{G}^{(1)}(\mathbf{r}_\mathrm{A},
 \mathbf{r}_\mathrm{A},\omega_{10})
 \!\cdot\!\mathbf{d}_{01},
\end{multline}
respectively.


\subsection{Dynamical theory}
\label{sec3.2}

In order to develop a dynamical theory of the CP force, which
particularly takes into account the line width of the
nonmonochromatic mode interacting with the atom, we start from the
operator-valued Lorentz force acting on the atom in electric-dipole
approximation \cite{0008},
\begin{align}
\label{eq44}
\hat{\mathbf{F}}&=\biggl\{\bm{\nabla}
 \bigl[\hat{\mathbf{d}}\cdot\hat{\mathbf{E}}(\mathbf{r})\bigr]
 +\frac{\mathrm{d}}{\mathrm{d}t}
 \bigl[\hat{\mathbf{d}}\times\hat{\mathbf{B}}(\mathbf{r})\bigr]
 \biggr\}_{\mathbf{r}=\hat{\mathbf{r}}_\mathrm{A}}
\nonumber\\
&\equiv\hat{\mathbf{F}}^\mathrm{el}
 +\hat{\mathbf{F}}^\mathrm{mag},
\end{align}
where we have separated the force into its electric and magnetic
parts. Note that while we have expressed the force in terms of
$\hat{\mathbf{E}}(\mathbf{r})$ which has the physical meaning of a
displacement field, Eq.~(\ref{eq44}) is also valid with the (physical)
electric field in place of $\hat{\mathbf{E}}(\mathbf{r})$
\cite{0008,0696}. For a two-level atom whose interaction with the
electromagnetic field is treated within the rotating-wave
approximation, we may employ Eqs.~(\ref{eq12}), (\ref{eq23}), and
(\ref{eq25}) to present $\hat{\mathbf{F}}^\mathrm{el}$ in the form
\begin{equation}
\label{eq45}
\hat{\mathbf{F}}^\mathrm{el}=-\hbar
\int_0^{\infty}\!\!\mathrm{d}\omega\,
 \Bigl[\bm{\nabla}g(\mathbf{r},\omega)
 \hat{a}(\mathrm{r},\omega)\hat{\sigma}^\dagger
 \Bigr]_{\mathbf{r}=\hat{\mathbf{r}}_\mathrm{A}}
 \!\!+\mathrm{H.c.},
\end{equation}
and by using Eq.~(\ref{2.31}), we may write
$\hat{\mathbf{F}}^\mathrm{mag}$ as
\begin{multline}
\label{F1}
\hat{\vect{F}}^\mathrm{mag}
 =\Biggl\{\frac{\dif}{\dif t}\sum_{\lambda=e,m}
 \int\dif^3r'\\
\times\int_0^\infty\frac{\dif\omega}{\mi\omega}\,\vect{d}_{10}
 \times\!
 \bigl[\vect{\nabla}\vprod\ten{G}_\lambda(\vect{r},\vect{r}',\omega)
 \bigr]\sprod\,\hat{\vect{f}}_\lambda(\vect{r}',\omega)
 \hat{\sigma}^\dagger\Biggr\}_{\vect{r}=\hat{\vect{r}}_\mathrm{A}}
\!\!\!\!\! +\mathrm{H.c.}
\end{multline}
For chosen atomic position
($\hat{\mathbf{r}}_\mathrm{A}\mapsto\mathbf{r}_\mathrm{A}$), the
dynamical CP force is just the average Lorentz force
\begin{align}
\label{eq46}
\mathbf{F}(\mathbf{r}_\mathrm{A},t)
& =\langle\psi(t)|\hat{\mathbf{F}}^\mathrm{el}|\psi(t)\rangle
 +\langle\psi(t)|\hat{\mathbf{F}}^\mathrm{mag}|\psi(t)\rangle
 \nonumber\\
&\equiv\mathbf{F}^\mathrm{el}(\mathbf{r}_\mathrm{A},t)
 +\mathbf{F}^\mathrm{mag}(\mathbf{r}_\mathrm{A},t),
\end{align}
where the state of the system, $|\psi(t)\rangle$,
evolves according to the Schr\"{o}dinger equation
\begin{equation}
\label{eq47}
\mi\hbar\frac{\partial}{\partial t}|\psi(t)\rangle
=\hat{H}|\psi(t)\rangle,
\end{equation}
with the Hamiltonian being given by Eq.~(\ref{eq1}), together with
Eqs.~(\ref{eq6}), (\ref{eq22}), and (\ref{eq34}).


\subsubsection{Quantum-state evolution}
\label{sec3.2.1}

In order to include in the evolution of the quantum state the loss
mechanisms associated with the line width of the nonmonochromatic mode
that strongly interacts with the atom as well as the interaction
of the atom with the residual, weakly interacting field, we
approximate $g^2(\mathbf{r},\omega)$, Eq.~(\ref{eq27}), as
\begin{equation}
\label{eq57}
g^2(\mathbf{r},\omega)
 =g_\nu^2(\mathbf{r},\omega_\nu)\,
 \frac{\gamma_\nu^2/4}{(\omega-\omega_\nu)^2+\gamma_\nu^2/4}\,
 +g^{\prime 2}(\mathbf{r},\omega),
\end{equation}
with the first and second term corresponding to the 
contributions from mode $\nu$ and the residual field, respectively. In
contrast to the drastic approximation that we have made in
Sec.~\ref{sec3.1} [recall Eq.~(\ref{eq35-1})], the width $\gamma_\nu$
of the mode is now finite, but it should still be small compared to
the separation of the mode from the neighboring ones, $\gamma_\nu$
$\ll$ $\delta\omega$. Furthermore, we have retained the residual field
continuum $g^{\prime 2}(\mathbf{r},\omega)$, which is assumed to be a
slowly varying function of $\omega$ in the vicinity of $\omega$ $\!=$
$\omega_\nu$ such that its effects on the dynamics is adequately
described by the Markov approximation, and
\begin{multline}
\label{eq57.1}
\!\!\!\!\!\int_{\omega_\nu-\delta\omega/2}
^{\omega_\nu+\delta\omega/2}
 \!\mathrm{d}\omega\,g^{\prime 2}(\mathbf{r},\omega)
\ll\!\int_{\omega_\nu-\delta\omega/2}^{\omega_\nu+\delta\omega/2}\!
 \mathrm{d}\omega\,\frac{g^2(\mathbf{r},\omega_\nu)\gamma_\nu^2/4}
 {(\omega-\omega_\nu)^2+\gamma_\nu^2/4}\\[.5ex]
={\textstyle\frac{1}{4}}\Omega^2_\mathrm{R}(\mathbf{r}_\mathrm{A})
\quad\mbox{for }\mathbf{r}=\mathbf{r}_\mathrm{A}.
\end{multline}
Under these assumptions, the natural generalization of the state
$|1_\nu\rangle$ defined in Eq.~(\ref{eq36.1}) reads
\begin{equation}
\label{eq57.2}
|1_\nu\rangle
 =\sqrt{\frac{2}{\pi\gamma_\nu}}\,
 \frac{1}{g(\mathbf{r}_\mathrm{A},\omega_\nu)}
 \int_{\omega_\nu-\delta\omega/2}^{\omega_\nu+\delta\omega/2}
 \mathrm{d}\omega\,g(\mathbf{r}_\mathrm{A},\omega)
 |\mathbf{r}_\mathrm{A},\omega\rangle.
\end{equation}
This state is normalized to unity, as can be easily seen by using
Eqs.~(\ref{eq32.2}) and (\ref{eq57.1}); it reduces to the
state~(\ref{eq36.1}) when setting $\gamma_\nu$ $\!\to$ $\!0$ and
$g^{\prime 2}(\mathbf{r},\omega)$ $\!\to$ $\!0$.

Let the system at initial time $t_0$ be prepared in a state
$|\psi(t_0)\rangle$ $\!=$ $\!|\theta\rangle$, where $|\theta\rangle$
is defined according to Eq.~(\ref{eq43.2}), but now with
$|1_\nu\rangle$ as given by Eq.~(\ref{eq57.1}). Within the
rotating-wave approximation, the state $|\psi_\theta(t)\rangle$ can
then be expanded as
\begin{equation}
\label{eq48}
|\psi(t)\rangle
 =\psi_1(t)|1\rangle|\{0\}\rangle
 +\int_0^\infty\mathrm{d}\omega\,\psi_0(\omega,t)
 |0\rangle|\mathbf{r}_\mathrm{A},\omega\rangle,
\end{equation}
where the coefficients $\psi_1(t)$ and $\psi_0(\omega,t)$ satisfy the
initial conditions
\begin{align}
\label{eq49}
&\psi_1(t_0)= \cos\theta,\\
\label{eq50}
&\psi_0(\omega,t_0)=\sqrt{\frac{2}{\pi\gamma_\nu}}\,
 \frac{g(\mathbf{r},\omega)}{g(\mathbf{r},\omega_\nu)}\,\sin\theta
 \bigl[\Theta(\omega-\omega_\nu+\delta\omega/2)\nonumber\\
&\hspace{25ex}-\Theta(\omega-\omega_\nu-\delta\omega/2)\bigr]
\end{align}
[$\Theta(x)$: unit step function].
We insert Eq.~(\ref{eq48}) in Eq.~(\ref{eq47}), make use of the
definition (\ref{eq30}), and recall the commutation relations
(\ref{eq28}) and (\ref{eq29}) to obtain the system of coupled
equations
\begin{align}
\label{eq52}
&\hspace{-1ex}\dot{\psi}_1(t)=-\mi\hbar^{-1}E_1\psi_1(t)
 -\mi\!\int_0^\infty\!
 \mathrm{d}\omega\,g(\mathbf{r}_\mathrm{A},\omega)
 \psi_0(\omega,t),\\[1ex]
\label{eq53}
&\hspace{-1ex}\dot{\psi}_0(\omega,t)=-\mi(\hbar^{-1}E_0+\omega)
 \psi_0(\omega,t)
 -\mi g(\mathbf{r}_\mathrm{A},\omega)\psi_1(t),
\end{align}
where $\psi_0(\omega,t)$ can be eliminated by formally solving
Eq.~(\ref{eq53}) [together with the initial condition (\ref{eq50})],
leading to
\begin{multline}
\label{eq54}
\psi_0(\omega,t)=\sqrt{\frac{2}{\pi\gamma_\nu}}\,
 \frac{g(\mathbf{r},\omega)}{g(\mathbf{r},\omega_\nu)}\,\sin\theta
 \me^{-\mi(E_0/\hbar+\omega)(t-t_0)}\\
 \bigl[\Theta(\omega-\omega_\nu+\delta\omega/2)
 -\Theta(\omega-\omega_\nu-\delta\omega/2)\bigr]\\
 -\mi g(\mathbf{r}_\mathrm{A},\omega)
\int_{t_0}^t\mathrm{d}\tau\,
 \me^{-\mi(E_0/\hbar+\omega)(t-\tau) }
 \psi_1(\tau).
\end{multline}
After substitution into Eq.~(\ref{eq52}), the frequency integral
resulting from the first term in Eq.~(\ref{eq54}) can be carried out
according to
\begin{align}
\label{eq56.1}
&-\sqrt{\frac{2}{\pi\gamma_\nu}}\,
 \frac{\mi\sin\theta}{g(\mathbf{r},\omega_\nu)}
\!\int_{\omega_\nu-\delta\omega/2}^{\omega_\nu+\delta\omega/2}
 \!\!\!\!
 \mathrm{d}\omega\,g^2(\mathbf{r}_\mathrm{A},\omega)
 \me^{-\mi(E_0/\hbar+\omega)(t-t_0)}\nonumber\\[.5ex]
&\quad\simeq
-{\textstyle\frac{1}{2}}\mi\Omega_\mathrm{R}(\mathbf{r}_\mathrm{A})
 \sin\theta\,\me^{[-\mi(E_0/\hbar+\omega_\nu)-\gamma_\nu/2](t-t_0)}
\nonumber\\[.5ex]
&\quad-\frac{
\mi
\Gamma'_1(\mathbf{r}_\mathrm{A})}
{\pi\Omega_\mathrm{R}(\mathbf{r}_\mathrm{A})}\,
  \me^{-\mi(E_0/\hbar+\omega_\nu)(t-t_0)}
 \delta_{1/\delta\omega}(t-t_0)
\end{align}
[$\delta_{1/\delta\omega}(t-t_0)$: coarse-grained delta function 
(time scale $1/\delta\omega$)], where we have used Eq.~(\ref{eq57})
together with the (approximately valid) relation \mbox{($\gamma_\nu$
$\!\ll$ $\!\delta\omega_\nu$)}
\begin{equation}
\label{eq36.5b}
\int_{\omega_\nu-\delta\omega_\nu/2}^{\omega_\nu+\delta\omega_\nu/2}
\mathrm{d}\omega\,\frac{\me^{-\mi\omega x}}
 {(\omega-\omega_\nu)^2+\gamma_\nu^2/4}
 =\frac{2\pi}{\gamma_\nu}\,\me^{-\mi\omega_\nu x-\gamma_\nu|x|/2}.
\end{equation}
The second term in Eq.~(\ref{eq56.1}), in which
$\Gamma'_1(\mathbf{r}_\mathrm{A})$, Eq.~(\ref{eq63}), is the
spontaneous decay rate of the upper atomic state due to the
weak coupling of the atom to the residual background field,  
is only non-vanishing near $t=t_0$, where its contribution is
negligible in comparison to that of the first term [recall
Eq.~(\ref{eq57.1})]. Hence Eqs.~(\ref{eq52}) and (\ref{eq54}) lead to
the following closed equation for $\psi_1(t)$:
\begin{multline}
\label{eq56}
\dot{\psi}_1(t)=-\mi\hbar^{-1}E_1\psi_1(t)\\
 -{\textstyle\frac{1}{2}}\mi\Omega_\mathrm{R}(\mathbf{r}_\mathrm{A})
 \sin\theta\,\me^{[-\mi(E_0/\hbar+\omega_\nu)-\gamma_\nu/2](t-t_0)}\\
 -\int_0^\infty\mathrm{d}\omega\,g^2(\mathbf{r}_\mathrm{A},\omega)
 \int_{t_0}^t\mathrm{d}\tau\,
 \me^{-\mi(E_0/\hbar+\omega)(t-\tau)}\psi_1(\tau).
\end{multline}

In order to solve this integro-differential equation, we make use of
Eq.~(\ref{eq57}) and treat the term $g^{\prime 2}(\mathbf{r},\omega)$,
which describes the effect of the residual field continuum, on the
basis of the Markov approximation. It can then be shown that (see
App.~\ref{AppA})
\begin{equation}
\label{eq58}
\psi_1(t)
 =\me^{[-\mi E_1/\hbar
 -\mi\delta\omega'_1(\mathbf{r}_\mathrm{A})
 -\Gamma'_1(\mathbf{r}_\mathrm{A})/2](t-t_0)}\phi_1(t),
\end{equation}
where $\phi_1(t)$ is the solution to the differential equation
\begin{multline}
\label{eq65}
\ddot{\phi}_1(t)
 +\bigl\{\mi\Delta(\mathbf{r}_\mathrm{A})+[\gamma_\nu
 -\Gamma'_1(\mathbf{r}_\mathrm{A})]/2\bigr\}
 \dot{\phi}_1(t)\\
+{\textstyle\frac{1}{4}}
 \Omega_\mathrm{R}^2(\mathbf{r}_\mathrm{A})\phi_1(t)=0
\end{multline}
together with the initial conditions
\begin{equation}
\label{eq66}
\phi_1(t_0)=\cos\theta,
 \quad\dot{\phi}_1(t_0)
 =-\frac{\mi}{2}\,\Omega_\mathrm{R}(\mathbf{r}_\mathrm{A})
 \sin\theta.
\end{equation}
Here,
\begin{multline}
\label{eq60}
\delta\omega'_1(\mathbf{r}_\mathrm{A})
=\frac{\mu_0}{\hbar\pi}\mathcal{P}
 \int_0^\infty\!\!\mathrm{d}\omega\,\omega^2\,
 \frac{\mathbf{d}_\mathrm{10}\!\cdot\!\mathrm{Im}\,
 \bm{G}^{(1)}(\mathbf{r}_\mathrm{A},\mathbf{r}_\mathrm{A},\omega)
 \!\cdot\!\mathbf{d}_\mathrm{01}}
 {\tilde{\omega}_{10}(\mathbf{r}_\mathrm{A})-\omega}\\
+\frac{\Omega^2_\mathrm{R}(\mathbf{r}_\mathrm{A})}{4}\,
 \frac{\Delta(\mathbf{r}_\mathrm{A})}
 {\Delta^2(\mathbf{r}_\mathrm{A})+\gamma_\nu^2/4}
\end{multline}
and
\begin{multline}
\label{eq63}
\Gamma'_1(\mathbf{r}_\mathrm{A})=\\
\frac{2\mu_0}{\hbar}\bigl[
 \tilde{\omega}_{10}(\mathbf{r}_\mathrm{A})\bigr]^2
 \mathbf{d}_\mathrm{01}\cdot\mathrm{Im}\,
 \bm{G}\bigl[\mathbf{r}_\mathrm{A},\mathbf{r}_\mathrm{A},
 \tilde{\omega}_{10}(\mathbf{r}_\mathrm{A})\bigr]
 \cdot\mathbf{d}_\mathrm{10}\\
-\frac{\Omega^2_\mathrm{R}(\mathbf{r}_\mathrm{A})}{4}\,
 \frac{\gamma_\nu}
 {\Delta^2(\mathbf{r}_\mathrm{A})+\gamma_\nu^2/4},
\end{multline}
respectively, are the shift and width of the upper level associated
with the residual field, and
\begin{equation}
\label{eq61}
\tilde{\omega}_{10}(\mathbf{r}_\mathrm{A})
 =\hbar^{-1}\bigl[\tilde{E}_1(\mathbf{r}_\mathrm{A})-E_0\bigr]
 =\omega_{10}+\delta\omega'_1(\mathbf{r}_\mathrm{A})
\end{equation}
and
\begin{equation}
\label{eq62}
\Delta(\mathbf{r}_\mathrm{A})
=\omega_\nu-\tilde{\omega}_{10}(\mathbf{r}_\mathrm{A})
=\Delta-\delta\omega'_1(\mathbf{r}_\mathrm{A})
\end{equation}
are the respective shifted atomic transition frequency and detuning.
Note that in Eq.~(\ref{eq60}), the vacuum Lamb shift contribution to
the level shift has been absorbed in the bare transition frequency
$\omega_{10}$ by making the replacement
\mbox{$\bm{G}\mapsto\bm{G}^{(1)}$}. Writing the general solution to
the differential equation (\ref{eq65}) in the form
\begin{equation}
\label{eq67}
\phi_1(t)=c_+(\mathbf{r}_\mathrm{A})
 \me^{\Omega_+(\mathbf{r}_\mathrm{A})(t-t_0)}
 +c_-(\mathbf{r}_\mathrm{A})
 \me^{\Omega_-(\mathbf{r}_\mathrm{A})(t-t_0)},
\end{equation}
we find that
\begin{multline}
\label{eq68}
\Omega_{\pm}(\mathbf{r}_\mathrm{A})=-{\textstyle\frac{1}{2}}
 \bigl\{
\mi\Delta(\mathbf{r}_\mathrm{A})
 \!+\![\gamma_\nu
 \!-\!\Gamma'_1(\mathbf{r}_\mathrm{A})]/2\bigr\}
\\[.5ex]
\mp{\textstyle\frac{1}{2}}
 \sqrt{\bigl\{\mi\Delta(\mathbf{r}_\mathrm{A})\!+\![\gamma_\nu
 \!-\!\Gamma'_1(\mathbf{r}_\mathrm{A})]/2\bigr\}^2
 -\Omega^2_\mathrm{R}(\mathbf{r}_\mathrm{A})}\,,
\end{multline}
and the initial conditions (\ref{eq66}) imply
\begin{equation}
\label{eq69}
c_\pm(\mathbf{r}_\mathrm{A})
 =\frac{\Omega_\mp(\mathbf{r}_\mathrm{A})\cos\theta
 +\frac{1}{2}\mi\Omega_\mathrm{R}(\mathbf{r}_\mathrm{A})
 \sin\theta}{\Omega_\mp(\mathbf{r}_\mathrm{A})
 -\Omega_\pm(\mathbf{r}_\mathrm{A})}\,.
\end{equation}


\subsubsection{Average Lorentz force}
\label{sec3.2.2}

Substituting Eqs.~(\ref{eq45}), (\ref{F1}), and (\ref{eq48}) into
Eq.~(\ref{eq46}), recalling Eq.~(\ref{eq30}), and using the
commutation relations (\ref{eq7}), (\ref{eq8}), (\ref{eq28}), and
(\ref{eq29}), we may express the average Lorentz force in terms
of $\psi_1(t)$ and $\psi_0(\omega,t)$ to obtain
\begin{multline}
\label{eq51}
\mathbf{F}^\mathrm{el}(\mathbf{r}_\mathrm{A},t)
 \equiv\mathbf{F}_\theta^\mathrm{el}(\mathbf{r}_\mathrm{A},t)=\\
- \hbar\!\int_0^{\infty}\!\!\!\mathrm{d}\omega\,
 \psi_1^\ast(t)\psi_0(\omega,t)\frac{
 \bigl[\bm{\nabla}g^{(1)}
 (\mathbf{r},\mathbf{r}_\mathrm{A},\omega)
 \bigr]_{\mathbf{r}=\mathbf{r}_\mathrm{A}}}
 {g(\mathbf{r}_\mathrm{A},\omega)}
 +\mathrm{C.c.}
\end{multline}
and
\begin{multline}
\label{F2}
\vect{F}^\mathrm{mag}(\vect{r}_\mathrm{A},t)
\equiv\vect{F}_\theta^\mathrm{mag}(\vect{r}_\mathrm{A},t)\\
=-\frac{\mu_0}{\mi\pi}\int_0^\infty\dif\omega\,\omega\,
 \frac{\dif}{\dif t}
 \bigl[\psi_1^\ast(t)\psi_0(\omega,t)\bigr]\\
 \times\frac{\vect{d}_{10}\vprod
 \bigl[\vect{\nabla}\vprod\mathrm{Im}\,
 \ten{G}^{(1)}(\vect{r},\vect{r}_\mathrm{A},\omega)
 \sprod\vect{d}_{01}
 \bigr]_{\vect{r}=\vect{r}_\mathrm{A}}}
 {g(\vect{r}_\mathrm{A},\omega)}\,
  +\mathrm{C.c.},
\end{multline}
where $g^{(1)}(\mathbf{r},\mathbf{r}',\omega)$ is defined according
to Eq.~(\ref{eq28.1}) with $\bm{G}(\mathbf{r},\mathbf{r}',\omega)$
being replaced with $\bm{G}^{(1)}(\mathbf{r},\mathbf{r}',\omega)$.
Note that the bulk part of the Green tensor,
$\bm{G}^{(0)}(\mathbf{r},\mathbf{r}',\omega)$,
does not contribute to the force. By making use of Eq.~(\ref{eq54}),
we can express $\vect{F}_\theta^\mathrm{el}(\vect{r}_\mathrm{A},t)$
and $\vect{F}_\theta^\mathrm{mag}(\vect{r}_\mathrm{A},t)$
entirely in terms of $\psi_1(t)$. For example,
$\vect{F}_\theta^\mathrm{el}(\vect{r}_\mathrm{A},t)$
can be given in the form
\begin{multline}
\label{eq55}
\mathbf{F}_\theta^\mathrm{el}(\mathbf{r}_\mathrm{A},t)
 =-\frac{\hbar\pi\gamma_\nu\sin\theta}
 {\Omega_\mathrm{R}(\mathbf{r}_\mathrm{A})}\,
 \me^{[-\mi(E_0/\hbar+\omega_\nu)-\gamma_\nu/2](t-t_0)}
 \psi_1^\ast(t)\\
\times\!
\bigl[\bm{\nabla}g^{(1)}
(\mathbf{r},\mathbf{r}_\mathrm{A},\omega_\nu)
 \bigr]_{\mathbf{r}=\mathbf{r}_\mathrm{A}}
 \!+\mi\hbar\!\int_0^{\infty}\!\!\mathrm{d}\omega\,
 \bigl[\bm{\nabla}g^{(1)}
 (\mathbf{r},\mathbf{r}_\mathrm{A},\omega)
 \bigr]_{\mathbf{r}=\mathbf{r}_\mathrm{A}}\\
\times\int_{t_0}^t\mathrm{d}\tau\,
 \me^{-\mi(E_0/\hbar+\omega)(t-\tau)}
 \psi_1^\ast(t)\psi_1(\tau)+\mathrm{C.c.},
\end{multline}
where, in close analogy to the derivation of Eq.~(\ref{eq56}), 
the first term has been obtained by means of the relation
\begin{equation}
\label{eq35b}
g(\mathbf{r},\mathbf{r}',\omega)
 =g_\nu(\mathbf{r},\mathbf{r}',\omega_\nu)\,
 \frac{\gamma_\nu^2/4}{(\omega-\omega_\nu)^2+\gamma_\nu^2/4}
 +g'(\mathbf{r},\mathbf{r}',\omega)
\end{equation}
[which reduces to Eq.~(\ref{eq57}) for $\mathbf{r}$ $\!=$
$\!\mathbf{r}'$], upon using Eq.~(\ref{eq36.5b}) and discarding the
term proportional to $\delta_{1/\delta\omega}(t-t_0)$, recall
Eq.~(\ref{eq56.1}).

Now we can make use of $\psi_1(t)$ as given by Eq.~(\ref{eq58})
together with Eq.~(\ref{eq67}). From Eq.~(\ref{eq55}) for
$\mathbf{F}_\theta^\mathrm{el}(\mathbf{r}_\mathrm{A},t)$
and from the analogous equation for
$\mathbf{F}_\theta^\mathrm{mag}(\mathbf{r}_\mathrm{A},t)$
we then obtain, after some algebra, the following expressions for the
electric and magnetic components of the resonant part of the force
observed in the case when the system is initially prepared in a state
of the type (\ref{eq43.2}):
\begin{multline}
\label{eq55b}
\mathbf{F}_\theta^\mathrm{el}(\mathbf{r}_\mathrm{A},t)\\
 =-\frac{\hbar\pi\gamma_\nu\sin\theta}
 {\Omega_\mathrm{R}(\mathbf{r}_\mathrm{A})}
 \bigl[\bm{\nabla}
 g^{(1)}(\mathbf{r},\mathbf{r}_\mathrm{A},\omega_\nu)
 \bigr]_{\mathbf{r}=\mathbf{r}_\mathrm{A}}
 q(\mathbf{r}_\mathrm{A},t-t_0)\\
+\hbar\int_0^{\infty}\mathrm{d}\omega\,
 \bigl[\bm{\nabla}
 g^{(1)}(\mathbf{r},\mathbf{r}_\mathrm{A},\omega)
 \bigr]_{\mathbf{r}=\mathbf{r}_\mathrm{A}}
 s(\mathbf{r}_\mathrm{A},\omega,t-t_0)+\mathrm{C.c.},
\end{multline}
\begin{multline}
\label{F3}
\vect{F}_\theta^\mathrm{mag}(\vect{r}_\mathrm{A},t)
 =\frac{\mi\mu_0\omega_\nu\gamma_\nu}
 {\Omega_\mathrm{R}(\vect{r}_\mathrm{A})}
  \sin\theta\,\frac{\dif}{\dif t}\,q(\vect{r}_\mathrm{A},t\!-\!t_0)\\
\times \vect{d}_{10}\vprod
 \bigl[\vect{\nabla}\vprod\mathrm{Im}\,
 \ten{G}^{(1)}(\vect{r},\vect{r}_\mathrm{A},\omega_\nu)\sprod
 \vect{d}_{01}\bigr]_{\vect{r}=\vect{r}_\mathrm{A}}\\
-\frac{\mi\mu_0}{\pi}\int_0^\infty\dif\omega\,\omega\,
 \vect{d}_{10}\vprod
 \bigl[\vect{\nabla}\vprod\mathrm{Im}\,
 \ten{G}^{(1)}
 (\vect{r},\vect{r}_\mathrm{A},\omega)\sprod\vect{d}_{01}
 \bigr]_{\vect{r}=\vect{r}_\mathrm{A}}\\
 \times\frac{\dif}{\dif t}\,
 s(\vect{r}_\mathrm{A},\omega,t-t_0)+\mathrm{C.c.},
\end{multline}
where
\begin{multline}
\label{eq70x1}
q(\mathbf{r}_\mathrm{A},t)
=\me^{\{-\mi\Delta(\mathbf{r}_\mathrm{A})
 -[\gamma_\nu+\Gamma'_1(\mathbf{r}_\mathrm{A})]/2\}t}\\
 \times\Bigl[c^\ast_+(\mathbf{r}_\mathrm{A})
 \me^{\Omega^\ast_+(\mathbf{r}_\mathrm{A})t}
 +c^\ast_-(\mathbf{r}_\mathrm{A})
 \me^{\Omega^\ast_-(\mathbf{r}_\mathrm{A})t}\Bigr]
\end{multline}
and
\begin{widetext}
\begin{align}
\label{eq70x2}
s(\mathbf{r}_\mathrm{A},\omega,t)
=&\,|c_+(\mathbf{r}_\mathrm{A})|^2\,
 \frac{\me^{[-\Gamma_1'(\mathbf{r}_\mathrm{A})
 +\Omega_+^\ast(\mathbf{r}_\mathrm{A})
 +\Omega_+(\mathbf{r}_\mathrm{A})]t}
 -\me^{\{\mi[\tilde{\omega}'_{10}(\mathbf{r}_\mathrm{A})-\omega]
 -\Gamma_1'(\mathbf{r}_\mathrm{A})/2
 +\Omega_+^\ast(\mathbf{r}_\mathrm{A})\}t} }
 {\omega-\tilde{\omega}'_{10}(\mathbf{r}_\mathrm{A})
 +\mi\Gamma_1'(\mathbf{r}_\mathrm{A})/2
 -\mi\Omega_+(\mathbf{r}_\mathrm{A})}
 \nonumber\\
&+c_+^\ast(\mathbf{r}_\mathrm{A}) c_-(\mathbf{r}_\mathrm{A})\,
 \frac{\me^{[-\Gamma_1'(\mathbf{r}_\mathrm{A})
 +\Omega_+^\ast(\mathbf{r}_\mathrm{A})
 +\Omega_-(\mathbf{r}_\mathrm{A})]t}
 -\me^{\{\mi[\tilde{\omega}'_{10}(\mathbf{r}_\mathrm{A})-\omega]
 -\Gamma_1'(\mathbf{r}_\mathrm{A})/2
 +\Omega_+^\ast(\mathbf{r}_\mathrm{A})\}t} }
 {\omega-\tilde{\omega}'_{10}(\mathbf{r}_\mathrm{A})
 +\mi\Gamma_1'(\mathbf{r}_\mathrm{A})/2
 -\mi\Omega_-(\mathbf{r}_\mathrm{A})}
 \nonumber\\
&+c_-^\ast(\mathbf{r}_\mathrm{A}) c_+(\mathbf{r}_\mathrm{A})\,
 \frac{\me^{[-\Gamma_1'(\mathbf{r}_\mathrm{A})
 +\Omega_-^\ast(\mathbf{r}_\mathrm{A})
 +\Omega_+(\mathbf{r}_\mathrm{A})]t}
 -\me^{\{\mi[\tilde{\omega}'_{10}(\mathbf{r}_\mathrm{A})-\omega]
 -\Gamma_1'(\mathbf{r}_\mathrm{A})/2
 +\Omega_-^\ast(\mathbf{r}_\mathrm{A})\}t} }
 {\omega-\tilde{\omega}'_{10}(\mathbf{r}_\mathrm{A})
 +\mi\Gamma_1'(\mathbf{r}_\mathrm{A})/2
-\mi\Omega_+(\mathbf{r}_\mathrm{A})}
 \nonumber\\
&+|c_-(\mathbf{r}_\mathrm{A})|^2\,
 \frac{\me^{[-\Gamma_1'(\mathbf{r}_\mathrm{A})
 +\Omega_-^\ast(\mathbf{r}_\mathrm{A})
 +\Omega_-(\mathbf{r}_\mathrm{A})]t}
 -\me^{\{\mi[\tilde{\omega}'_{10}(\mathbf{r}_\mathrm{A})-\omega]
 -\Gamma_1'(\mathbf{r}_\mathrm{A})/2
 +\Omega_-^\ast(\mathbf{r}_\mathrm{A})\}t} }
 {\omega-\tilde{\omega}'_{10}(\mathbf{r}_\mathrm{A})
 +\mi\Gamma_1'(\mathbf{r}_\mathrm{A})/2
 -\mi\Omega_-(\mathbf{r}_\mathrm{A})}\,.
\end{align}
\end{widetext}


\subsubsection{Weak and strong coupling limits}
\label{sec3.2.3}

In order to gain deeper insight into the results, let us consider the
limiting cases of weak and strong atom--field coupling in more detail.
To make contact with earlier results, we begin with the weak-coupling
regime.


\paragraph{Weak atom--field coupling.}

Weak atom--field coupling is realized if the width $\gamma_\nu$ of
the nonmonochromatic mode is sufficiently large,
\begin{equation}
\label{eq71}
\gamma_\nu\gg 2\Omega_\mathrm{R}(\mathbf{r}_\mathrm{A}),
\end{equation}
and/or the detuning $\Delta(\mathbf{r}_\mathrm{A})$ is sufficiently
large,
\begin{equation}
\label{eq72}
|\Delta(\mathbf{r}_\mathrm{A})|\gg
2\Omega^2_\mathrm{R}(\mathbf{r}_\mathrm{A})/\gamma_\nu.
\end{equation}
In both cases, the first term under the square root in
Eq.~(\ref{eq68}) is much larger than the second one and a Taylor
expansion yields the approximations
\begin{align}
\label{eq73}
&\Omega_{\pm}(\mathbf{r}_\mathrm{A})
= \nonumber\\[.5ex]
&\begin{cases}
-\mi\Delta(\mathbf{r}_\mathrm{A})-[\gamma_\nu
 -\Gamma'_1(\mathbf{r}_\mathrm{A})]/2\\[1ex]
{\displaystyle
 \frac{\mi\Omega^2_\mathrm{R}(\mathbf{r}_\mathrm{A})}{4}
 \,\frac{\Delta(\mathbf{r}_\mathrm{A})}
 {\Delta^2(\mathbf{r}_\mathrm{A})\!+\!\gamma_\nu^2/4}
 -\frac{\Omega^2_\mathrm{R}(\mathbf{r}_\mathrm{A})}{8}\,
 \frac{\gamma_\nu}{\Delta^2(\mathbf{r}_\mathrm{A})
 \!+\!\gamma_\nu^2/4}}\,.
\end{cases}
\end{align}
For a system initially prepared in the state $|1\rangle|\{0\}\rangle$,
i.e., $\theta$ $\!=$ $\!0$ [recall Eq.~(\ref{eq43.3})],
Eq.~(\ref{eq69}) approximates to
\begin{equation}
\label{eq77}
c_+(\mathbf{r}_\mathrm{A})=0, \quad
c_-(\mathbf{r}_\mathrm{A})=1
\end{equation}
because $|\Omega_+(\mathbf{r}_\mathrm{A})|$ $\!\gg$
$\!|\Omega_-(\mathbf{r}_\mathrm{A})|$. Substituting Eqs.~(\ref{eq58}),
(\ref{eq67}), (\ref{eq73}), and (\ref{eq77}) into Eq.~(\ref{eq48})
and recalling Eqs.~(\ref{eq60}) and (\ref{eq63}), we find that in the
weak-coupling limit the temporal evolution of the system initially
prepared in the state $|1\rangle|\{0\}\rangle$ may be given by
\begin{align}
\label{eq78}
|\psi(t)\rangle
& =\me^{[-\mi E_1/\hbar
 -\mi\delta\omega'_1(\mathbf{r}_\mathrm{A})
 -\Gamma'_1(\mathbf{r}_\mathrm{A})/2
 +\Omega_-(\mathbf{r}_\mathrm{A})](t-t_0)}\,
 |1\rangle|\{0\}\rangle\nonumber\\
& =\me^{[-\mi\tilde{E}_1(\mathbf{r}_\mathrm{A})/\hbar
 -\Gamma_1(\mathbf{r}_\mathrm{A})/2](t-t_0)}
 |1\rangle|\{0\}\rangle.
\end{align}
Here, the shift and width of the upper atomic level are given by
\begin{equation}
\label{eq80}
\delta\omega_1(\mathbf{r}_\mathrm{A})
=\frac{\mu_0}{\hbar\pi}\mathcal{P}
 \int_0^\infty\mathrm{d}\omega\,\omega^2
 \frac{\mathbf{d}_\mathrm{10}\!\cdot\!\mathrm{Im}\,
 \bm{G}^{(1)}(\mathbf{r}_\mathrm{A},\mathbf{r}_\mathrm{A},\omega)
 \!\cdot\!\mathbf{d}_\mathrm{01}}
 {\tilde{\omega}_\mathrm{10}(\mathbf{r}_\mathrm{A})-\omega}
\end{equation}
and
\begin{equation}
\label{eq81}
\Gamma_1(\mathbf{r}_\mathrm{A})=
 \frac{2\mu_0}{\hbar}
 \bigl[\tilde{\omega}_{10}(\mathbf{r}_\mathrm{A})\bigr]^2
 \mathbf{d}_\mathrm{10}\!\cdot\!\mathrm{Im}\,
 \bm{G}\bigl[\mathbf{r}_\mathrm{A},\mathbf{r}_\mathrm{A},
 \tilde{\omega}_{10}(\mathbf{r}_\mathrm{A})\bigr]
 \!\cdot\!\mathbf{d}_\mathrm{01},
\end{equation}
respectively, and the shifted atomic transition frequency
$\tilde{\omega}_{10}(\mathbf{r}_\mathrm{A})$ reads
\begin{equation}
\label{eq79}
\tilde{\omega}_{10}(\mathbf{r}_\mathrm{A})
 =\hbar^{-1}\bigl[\tilde{E}_1(\mathbf{r}_\mathrm{A})-E_0\bigr]
 =\omega_{10}+\delta\omega_1(\mathbf{r}_\mathrm{A})
\end{equation}
in place of Eq.~(\ref{eq61}). Note that in the weak-coupling limit the
two terms in the decomposition (\ref{eq57}) can be treated on an equal
footing, so that the decomposition becomes superfluous.

The associated force in the weak-coupling limit can most conveniently
be derived by returning to Eq.~(\ref{eq55}) [and the analogous
equation for
$\mathbf{F}_\theta^\mathrm{mag}(\mathbf{r}_\mathrm{A},t)$]
and making therein use of  Eq.~(\ref{eq78}) in the form
\begin{align}
\label{eq78b}
\psi_1(t)
 =\me^{[-\mi\tilde{E}_1(\mathbf{r}_\mathrm{A})/\hbar
 -\Gamma_1(\mathbf{r}_\mathrm{A})/2](t-\tau)}\,\psi_1(\tau).
\end{align}
Evaluating the time integral in the spirit of the Markov
approximation by setting $|\psi_1(\tau)|^2\mapsto|\psi_1(t)|^2$
and letting the lower integration limit tend to
$-\infty$, one finds, on recalling Eq.~(\ref{eq28.1}), that
$\mathbf{F}_{1\{0\}}^\mathrm{mag}(\mathbf{r}_\mathrm{A},t)$
$\!=$ $\!0$ and hence,
\begin{equation}
\label{eq82}
\mathbf{F}_{1\{0\}}(\mathbf{r}_\mathrm{A},t) =
\mathbf{F}_{1\{0\}}^\mathrm{el}(\mathbf{r}_\mathrm{A},t)
=\me^{-\Gamma_1(\mathbf{r}_\mathrm{A})(t-t_0)}
\mathbf{F}_{1\{0\}}(\mathbf{r}_\mathrm{A}),
\end{equation}
where
\begin{multline}
\label{eq82b}
\mathbf{F}_{1\{0\}}(\mathbf{r}_\mathrm{A})=\\
\frac{\mu_0}{\pi}\!\!
\int_0^\infty\!\!\mathrm{d}\omega\,\omega^2
\frac{\bigl[\bm{\nabla}
 \mathbf{d}_{10}\!\cdot\!\mathrm{Im}\,
 \bm{G}^{(1)}(\mathbf{r},\mathbf{r}_\mathrm{A},\omega)
 \!\cdot\!\mathbf{d}_{01}\bigr]_{\mathbf{r}=\mathbf{r}_\mathrm{A}}}
 {\omega-\tilde{\omega}_{10}(\mathbf{r}_\mathrm{A})
 -\mi\Gamma_1(\mathbf{r}_\mathrm{A})/2}
 +\mathrm{C.c.}\\
\simeq\mu_0\Omega^2_{10}(\mathbf{r}_{A})
 \bigl\{\bm{\nabla}\mathbf{d}_{10}\cdot
 \bm{G}^{(1)}[\mathbf{r},\mathbf{r}_\mathrm{A},
 \Omega_{10}(\mathbf{r}_\mathrm{A})]
 \cdot\mathbf{d}_{01}
 \bigr\}_{\mathbf{r}=\mathbf{r}_\mathrm{A}}\\
+\mathrm{C.c.}
\end{multline}
with
\begin{equation}
\label{eq82c}
\Omega_{10}(\mathbf{r}_\mathrm{A})
=\tilde{\omega}_{10}(\mathbf{r}_\mathrm{A})
 +\mi\Gamma_1(\mathbf{r}_\mathrm{A})/2.
\end{equation}
As expected, Eq.~(\ref{eq82}) agrees with the resonant contribution to
the force as derived in Refs.~\cite{0008,0012}, in the special case of
a two level atom that is initially prepared in the upper state.


\paragraph{Strong atom--field coupling.}

Having thus established that the general
expression~(\ref{eq55b})--(\ref{eq70x2}) reproduces earlier
results in the weak-coupling limit, we now turn our attention to the
strong-coupling regime. Strong atom--field coupling is realized if
both the spectral width $\gamma_\nu$ of mode $\nu$ and the width
$\Gamma'_1(\mathbf{r}_\mathrm{A})$ of the upper atomic level, which is
associated with the residual field, are sufficiently small,
\begin{equation}
\label{eq85}
\gamma_\nu,\Gamma'_1(\mathbf{r}_\mathrm{A})\ll
2\Omega_\mathrm{R}(\mathbf{r}_\mathrm{A}),
\end{equation}
and its mid-frequency $\omega_\nu$ is sufficiently close to the atomic
transition frequency $\tilde{\omega}_{10}(\mathbf{r}_\mathrm{A})$,
\begin{equation}
\label{eq86}
|\Delta(\mathbf{r}_\mathrm{A})|\ll
2\Omega^2_\mathrm{R}(\mathbf{r}_\mathrm{A})/\gamma_\nu.
\end{equation}

Even when the coupling is at least moderately strong such that
$\gamma_\nu$, $\!\Gamma'_1(\mathbf{r}_\mathrm{A})$
$\!\le$ $\!2\Omega_\mathrm{R}(\mathbf{r}_\mathrm{A})$ and the
inequality (\ref{eq86}) holds [note that in addition,
$|\Delta(\mathbf{r}_\mathrm{A})|\Gamma'_1(\mathbf{r}_\mathrm{A})$
$\!\ll$ $\!2\Omega^2_\mathrm{R}(\mathbf{r}_\mathrm{A})$ is
automatically fulfilled as a direct consequence of Eqs.~(\ref{eq57.1})
and (\ref{eq63})], then the real part of the square root in
Eq.~(\ref{eq68}) becomes negligibly small so that
\begin{equation}
\label{eq87}
\Omega_{\pm}(\mathbf{r}_\mathrm{A})=-{\textstyle\frac{1}{2}}
 \bigl\{\mi\Delta(\mathbf{r}_\mathrm{A})
 \!+\![\gamma_\nu
 \!-\!\Gamma'_1(\mathbf{r}_\mathrm{A})]/2\bigr\}
 \mp{\textstyle\frac{1}{2}}\mi
 \Omega(\mathbf{r}_\mathrm{A})
\end{equation}
is approximately valid, where $\Omega(\mathbf{r}_\mathrm{A})$ is now
given by
\begin{equation}
\label{eq88}
\Omega(\mathbf{r}_\mathrm{A})
=\sqrt{\Omega_\mathrm{R}^2(\mathbf{r}_\mathrm{A})
 +\Delta^2(\mathbf{r}_\mathrm{A})
 -\bigl[\gamma_\nu
 \!-\!\Gamma'_1(\mathbf{r}_\mathrm{A})\bigr]^2/4}
\end{equation}
in place of Eq.~(\ref{eq43c}). Accordingly, $\psi_1(t)$
[Eq.~(\ref{eq58}) together with Eq.~(\ref{eq67})] approximately
reads
\begin{multline}
\label{eq92}
\psi_1(t)
=e^{-\gamma(\mathbf{r}_\mathrm{A})(t-t_0)/2}\Bigl[
 c_+(\mathbf{r}_\mathrm{A})
 \me^{-\mi E_+(\mathbf{r}_\mathrm{A})(t-t_0)/\hbar}\\
+c_-(\mathbf{r}_\mathrm{A})
 \me^{-\mi E_-(\mathbf{r}_\mathrm{A})(t-t_0)/\hbar}\Bigr],
\end{multline}
where
\begin{equation}
\label{eq94}
E_\pm(\mathbf{r}_\mathrm{A})
=\frac{E_0+E_1+\hbar\delta\omega'_1(\mathbf{r}_\mathrm{A})
+\hbar\omega_\nu}{2}
 \pm\frac{\hbar}{2}\,\Omega(\mathbf{r}_\mathrm{A})
\end{equation}
are the eigenenergies of the system in place of Eq.~(\ref{eq40}),
\begin{equation}
\label{eq68-1}
\gamma(\mathbf{r}_\mathrm{A})
 ={\textstyle\frac{1}{2}}[\gamma_\nu
 +\Gamma'_1(\mathbf{r}_\mathrm{A})]
\end{equation}
is the total damping rate and the coefficients
$c_\pm(\mathbf{r}_\mathrm{A})$ are given by Eq.~(\ref{eq69}) with
$\Omega_{\pm}(\mathbf{r}_\mathrm{A})$ from Eq.~(\ref{eq87}). Note that
from Eq.~(\ref{eq68-1}) [together with Eq.~(\ref{eq63})] and
Eq.~(\ref{eq60}) it follows that
\begin{equation}
\label{eq93}
\gamma(\mathbf{r}_\mathrm{A})
\simeq\begin{cases}
\gamma_\nu/2,
 \quad |\Delta(\mathbf{r}_\mathrm{A})|\ll\gamma_\nu/2,\\[1ex]
\Gamma_1(\mathbf{r}_\mathrm{A})/2,
\quad \gamma_\nu/2\ll|\Delta(\mathbf{r}_\mathrm{A})|
 \ll 2\Omega^2_\mathrm{R}(\mathbf{r}_\mathrm{A})/\gamma_\nu
\end{cases}
\end{equation}
and
\begin{equation}
\label{eq95}
\delta\omega'_1(\mathbf{r}_\mathrm{A})
\simeq\begin{cases}
\Omega^2_\mathrm{R}(\mathbf{r}_\mathrm{A})
 \Delta(\mathbf{r}_\mathrm{A})/\gamma_\nu^2,
 \quad|\Delta(\mathbf{r}_\mathrm{A})|\ll\gamma_\nu/2,\\[1ex]
\delta\omega_1(\mathbf{r}_\mathrm{A}),
 \quad\gamma_\nu/2\ll|\Delta(\mathbf{r}_\mathrm{A})|
 \ll 2\Omega^2_\mathrm{R}(\mathbf{r}_\mathrm{A})/\gamma_\nu
\end{cases}
\end{equation}
[recall Eqs.~(\ref{eq80}) and (\ref{eq81})].

We substitute Eq.~(\ref{eq87}) into Eqs.~(\ref{eq55b}) and (\ref{F3})
and calculate the frequency integrals by means of Eq.~(\ref{eq35b}).
The residual-field term $g'(\mathbf{r},\mathbf{r}',\omega)$ can
be pulled out of the integration and does hence not contribute to the
resonant force [since the only remaining poles giving rise to resonant
terms are then those of $s(\mathbf{r}_\mathrm{A},\omega,t-t_0)$, at
which the respective numerators vanish, cf. Eq.~(\ref{eq70x2})]. The
nonvanishing contribution due to mode $\nu$ can be evaluated in the
spirit of Eq.~(\ref{eq36.5b}), and after a tedious, but
straightforward calculation, we arrive at
\begin{multline}
\label{eq95.2}
\mathbf{F}_\theta^\mathrm{el}(\mathbf{r}_\mathrm{A},t)
 =-\hbar\pi\gamma_\nu
 \bigl[\bm{\nabla}g
^{(1)}
 (\mathbf{r},\mathbf{r}_\mathrm{A},\omega_\nu)
 \bigr]_{\mathbf{r}=\mathbf{r}_\mathrm{A}}
 \me^{-\gamma(\mathbf{r}_\mathrm{A})(t-t_0)}\\
 \times\Biggl[
 \frac{|c_+(\mathbf{r}_\mathrm{A})|^2
 +c_-^\ast(\mathbf{r}_\mathrm{A})c_+(\mathbf{r}_\mathrm{A})
 \me^{-\mi\Omega(\mathbf{r}_\mathrm{A})(t-t_0)}}
 {\Delta(\mathbf{r}_\mathrm{A})-\Omega(\mathbf{r}_\mathrm{A})
 -\mi\bigl[\gamma_\nu
 \!-\!\Gamma'_1(\mathbf{r}_\mathrm{A})\bigr]/2}\\
 +\frac{|c_-(\mathbf{r}_\mathrm{A})|^2
 +c_+^\ast(\mathbf{r}_\mathrm{A})c_-(\mathbf{r}_\mathrm{A})
 \me^{\mi\Omega(\mathbf{r}_\mathrm{A})(t-t_0)}}
 {\Delta(\mathbf{r}_\mathrm{A})+\Omega(\mathbf{r}_\mathrm{A})
 -\mi\bigl[\gamma_\nu
 \!-\!\Gamma'_1(\mathbf{r}_\mathrm{A})\bigr]/2}\Biggr]
 +\mathrm{C.c.}
\end{multline}
and
\begin{multline}
\label{F4}
\vect{F}_\theta^\mathrm{mag}(\vect{r}_\mathrm{A},t)
 =\mu_0\omega_\nu\gamma_\nu
 \Omega_\mathrm{R}(\vect{r}_\mathrm{A})\\
\times\vect{d}_{10}\vprod
 \bigl[\vect{\nabla}\vprod\mathrm{Im}\,
 \ten{G}^{(1)}(\vect{r},\vect{r}_\mathrm{A},\omega_\nu)
 \sprod\vect{d}_{01}
 \bigr]_{\vect{r}=\vect{r}_\mathrm{A}}
 \me^{-\gamma(\vect{r}_\mathrm{A})(t-t_0)}\\
 \Biggl[\frac{c_-^\ast(\vect{r}_\mathrm{A})
 c_+(\vect{r}_\mathrm{A})
 \me^{-\mi\Omega(\vect{r}_\mathrm{A})(t-t_0)}}
 {\Delta(\vect{r}_\mathrm{A})-\Omega(\vect{r}_\mathrm{A})
 -\mi\bigl[\gamma_\nu
 -\Gamma'_1(\vect{r}_\mathrm{A})\bigr]/2}\\
 -\frac{c_+^\ast(\vect{r}_\mathrm{A})
 c_-(\vect{r}_\mathrm{A})
 \me^{\mi\Omega(\vect{r}_\mathrm{A})(t-t_0)}}
 {\Delta(\vect{r}_\mathrm{A})
 +\Omega(\vect{r}_\mathrm{A})
 -\mi\bigl[\gamma_\nu
 -\Gamma'_1(\vect{r}_\mathrm{A})\bigr]/2}\Biggr]
  +\mathrm{C.c.}
\end{multline}

In order to compare with the (purely electric) weak-coupling
result (\ref{eq82})--(\ref{eq82c}), let us consider the case
where the system is initially prepared in the state
$|1\rangle|\{0\}\rangle$ [i.e., $\theta$ $\!=$ $\!0$, recall
Eq.~(\ref{eq43.3})] and examine the electric part of the force in more
detail. Noting that the coefficients $c_\pm(\mathbf{r}_\mathrm{A})$,
Eq.~(\ref{eq69}), reduce to
\begin{multline}
\label{eq69c}
c_\pm(\mathbf{r}_\mathrm{A})
 =\frac{\Omega(\mathbf{r}_\mathrm{A})
 \mp\Delta(\mathbf{r}_\mathrm{A})\pm\mi
 \bigl[\gamma_\nu
 \!-\!\Gamma'_1(\mathbf{r}_\mathrm{A})\bigr]/2}
 {2\Omega(\mathbf{r}_\mathrm{A})}\,,
\end{multline}
so that for real dipole matrix elements where
\begin{align}
\label{eq95x}
\bm{\nabla}g^{(1)}(\mathbf{r},\mathbf{r}_\mathrm{A},\omega_\nu)
 |_{\mathbf{r}=\mathbf{r}_\mathrm{A}}
&=\bm{\nabla}g^{(1)\ast}(\mathbf{r},\mathbf{r}_\mathrm{A},\omega_\nu)
 |_{\mathbf{r}=\mathbf{r}_\mathrm{A}}\nonumber\\
&=g(\mathbf{r}_\mathrm{A},\omega_\nu)
 \bm{\nabla}_{\!\!\mathrm{A}}g(\mathbf{r}_\mathrm{A},\omega_\nu)
\end{align}
[recall Eqs.~(\ref{eq19}), (\ref{eq27}), and (\ref{eq28.1})], from
Eq.~(\ref{eq95.2}) it follows that
\begin{multline}
\label{eq95.3}
\mathbf{F}_{1\{0\}}^\mathrm{el}(\mathbf{r}_\mathrm{A},t)
 =\me^{-\gamma(\mathbf{r}_\mathrm{A})(t-t_0)}
 \sin^2[\Omega(\mathbf{r}_\mathrm{A})(t-t_0)/2]\\
\times\frac{\hbar\pi\gamma_\nu
 \bigl[\bm{\nabla}g^{(1)}(\mathbf{r},\mathbf{r}_\mathrm{A},\omega_\nu)
 \bigr]_{\mathbf{r}=\mathbf{r}_\mathrm{A}}}
 {\Delta(\mathbf{r}_\mathrm{A})
 +\mi\bigl[\gamma_\nu
 \!-\!\Gamma'_1(\mathbf{r}_\mathrm{A})\bigr]/2}\\
\times\frac{\Delta^2(\mathbf{r}_\mathrm{A})
 -[\gamma_\nu
 \!-\!\Gamma'_1(\mathbf{r}_\mathrm{A})]^2/4}
 {\Delta^2(\mathbf{r}_\mathrm{A})
 -[\gamma_\nu
 \!-\!\Gamma'_1(\mathbf{r}_\mathrm{A})]^2/4
 +\Omega_\mathrm{R}^2(\mathbf{r}_\mathrm{A})}
 +\mathrm{C.c.}
\end{multline}
Since, within the approximation scheme used [Eq.~(\ref{eq35b})], we
may set
\begin{equation}
\label{eq95.4}
\frac{2\mu_0}{\hbar\pi}\,\omega^2
 \mathbf{d}_{10}\!\cdot\!\bm{G}^{(1)}
(\mathbf{r},\mathbf{r}',\omega)
 \!\cdot\!\mathbf{d}_{01}
=\frac{\gamma_\nu g^{(1)}(\mathbf{r},\mathbf{r}',\omega_\nu)}
 {\omega_\nu-\omega+\mi\gamma_\nu/2}\,,
\end{equation}
we may rewrite Eq.~(\ref{eq95.3}) as
\begin{multline}
\label{eq95.5}
\mathbf{F}_{1\{0\}}^\mathrm{el}(\mathbf{r}_\mathrm{A},t)
 =2\me^{-\gamma(\mathbf{r}_\mathrm{A})(t-t_0)}
 \sin^2[\Omega(\mathbf{r}_\mathrm{A})(t-t_0)/2]
\\\times
C(\mathbf{r}_\mathrm{A})
\mathbf{F}_{1\{0\}}(\mathbf{r}_\mathrm{A}),
\end{multline}
where
\begin{equation}
\label{eq95.7}
\!C(\mathbf{r}_\mathrm{A})=
\frac{\Delta^2(\mathbf{r}_\mathrm{A})
 -[\gamma_\nu
 \!-\!\Gamma'_1(\mathbf{r}_\mathrm{A})/2]^2/4}
 {\Delta^2(\mathbf{r}_\mathrm{A})
 -[\gamma_\nu
 \!-\!\Gamma'_1(\mathbf{r}_\mathrm{A})/2]^2/4
 +\Omega_\mathrm{R}^2(\mathbf{r}_\mathrm{A})}\,,
\end{equation}
and $\mathbf{F}_{1\{0\}}(\mathbf{r}_\mathrm{A})$ is given by
Eq.~(\ref{eq82b}) with
\begin{equation}
\label{eq95.6}
\Omega_{10}(\mathbf{r}_{A})
=\tilde{\omega}'_{10}(\mathbf{r}_\mathrm{A})
 +\mi\Gamma'_1(\mathbf{r}_\mathrm{A})/2
\end{equation}
in place of Eq.~(\ref{eq82c}). Note that
\begin{equation}
\label{eq95.7-1}
C(\mathbf{r}_\mathrm{A})
\simeq\begin{cases}
 {\displaystyle\frac{\gamma_\nu^2/4}{\gamma_\nu^2/4
 \!-\!\Omega_\mathrm{R}^2(\mathbf{r}_\mathrm{A})}\,,}
 \quad|\Delta(\mathbf{r}_\mathrm{A})|\ll\gamma_\nu/2,\\[2ex]
 {\displaystyle\frac{\Delta^2(\mathbf{r}_\mathrm{A})}
 {\Delta^2(\mathbf{r}_\mathrm{A})
 \!+\!\Omega_\mathrm{R}^2(\mathbf{r}_\mathrm{A})}\,,} \\[1ex]
 \hspace{4ex}
 \gamma_\nu/2\!\ll\!|\Delta(\mathbf{r}_\mathrm{A})|
 \!\ll\! 2\Omega^2_\mathrm{R}(\mathbf{r}_\mathrm{A})/\gamma_\nu.
\end{cases}
\end{equation}

Equations (\ref{eq95.5})--(\ref{eq95.7-1}) reveal that the electric
part of the force for moderately strong to strong atom--field coupling
differs from the respective weak-coupling result
(\ref{eq82})--(\ref{eq82c}) in several respects. Firstly,
the strength of the force is modified by the correction factor
$C(\mathbf{r}_\mathrm{A})$. Secondly, and most strikingly, the time
dependence of the force is given not by a simple exponential decay,
but by damped Rabi oscillations with frequency
$\Omega(\mathbf{r}_\mathrm{A})$ and damping rate
$\gamma(\mathbf{r}_\mathrm{A})$---oscillations that are well
pronounced when the inequalities~(\ref{eq85}) hold. Note that only
the magnitude, and not the sign of the force is oscillating, with the
sign being determined by the sign of the detuning
$\Delta(\mathbf{r}_\mathrm{A})$, as for the force in the case of weak
atom--field coupling. As seen from Eq.~(\ref{eq93}), for small
detuning, $|\Delta(\mathbf{r}_\mathrm{A})|$ $\!\ll$ $\!\gamma_\nu/2$,
the damping is dominated by the radiative and non-radiative losses the
quasi-monochromatic mode suffers from, while radiative and
non-radiative spontaneous decay of the upper atomic state is the
dominant loss mechanism for large detuning, $\gamma_\nu/2$ $\!\ll$
$\!|\Delta(\mathbf{r}_\mathrm{A})|$ $\!\ll$
$\!2\Omega^2_\mathrm{R}(\mathbf{r}_\mathrm{A})/\gamma_\nu$, where in
both limits the damping of the force is characterized by one-half the
respective damping rates, $\gamma_\nu$ and
$\Gamma_1(\mathbf{r}_\mathrm{A})$, respectively. This can easily be
understood from the fact that the force depends on the product of the
electric-dipole moment of the atom and the electric-field strength of
the quasi-monochromatic mode, whose damping rates are
$\Gamma_1(\mathbf{r}_\mathrm{A})/2$ and $\gamma_\nu/2$, respectively.

Let us return to the case of arbitrary initial preparation of the
system. In the limit of strong atom--field coupling when the
inequalities (\ref{eq85}) and (\ref{eq86}) are fulfilled,
Eq.~(\ref{eq88}) simplifies to
\begin{equation}
\label{eq88b}
\Omega(\mathbf{r}_\mathrm{A})
=\sqrt{\Omega_\mathrm{R}^2(\mathbf{r}_\mathrm{A})
 +\Delta^2(\mathbf{r}_\mathrm{A})}\,.
\end{equation}
Similarly, after introducing the coupling angle
\begin{equation}
\label{eq89}
\tan[2\theta_c(\mathbf{r}_\mathrm{A})]
=-\frac{\Omega_\mathrm{R}(\mathbf{r}_\mathrm{A})}
 {\Delta(\mathbf{r}_\mathrm{A})},
 \quad\theta_c(\mathbf{r}_\mathrm{A})\in[0,\pi/2]
\end{equation}
[which replaces Eq.~(\ref{eq41})] and using the identities
(\ref{eq41.1}) and (\ref{eq73.16}), we may write the coefficients
$c_\pm(\mathbf{r}_\mathrm{A})$, Eq.~(\ref{eq69}), as
\begin{align}
\label{eq90}
c_+(\mathbf{r}_\mathrm{A})
=&\cos^2[\theta_c(\mathbf{r}_\mathrm{A})]\cos\theta
 +\sin[\theta_c(\mathbf{r}_\mathrm{A})]
 \cos[\theta_c(\mathbf{r}_\mathrm{A})]\sin\theta\nonumber\\
=&\cos[\theta_c(\mathbf{r}_\mathrm{A})]
\cos[\theta-\theta_c(\mathbf{r}_\mathrm{A})],\\[1ex]
\label{eq91}
c_-(\mathbf{r}_\mathrm{A})
=&\sin^2[\theta_c(\mathbf{r}_\mathrm{A})]\cos\theta
 -\sin[\theta_c(\mathbf{r}_\mathrm{A})]
 \cos[\theta_c(\mathbf{r}_\mathrm{A})]\sin\theta\nonumber\\
=&-\sin[\theta_c(\mathbf{r}_\mathrm{A})]
 \sin[\theta-\theta_c(\mathbf{r}_\mathrm{A})].
\end{align}
Equation (\ref{eq95.2}) for the electric part of the force then reads
\begin{multline}
\label{eq95.2b}
\mathbf{F}_\theta^\mathrm{el}(\mathbf{r}_\mathrm{A},t)=\\
 -\frac{\hbar\pi\gamma_\nu
 \bigl[\bm{\nabla}g^{(1)}
 (\mathbf{r},\mathbf{r}_\mathrm{A},\omega_\nu)
 \bigr]_{\mathbf{r}=\mathbf{r}_\mathrm{A}}}
 {2\Omega_\mathrm{R}(\mathbf{r}_\mathrm{A})}\,
 \me^{-\gamma(\mathbf{r}_\mathrm{A})(t-t_0)}
\\
\times\Bigl(\sin[2\theta_c(\mathbf{r}_\mathrm{A})]
 \cos\bigl\{2[\theta\!-\!\theta_c(\mathbf{r}_\mathrm{A})]\bigr\}
 +\sin\bigl\{2[\theta\!-\!\theta_c(\mathbf{r}_\mathrm{A})]\bigr\}
\\
+\!\Bigl\{\!\cos^2[\theta_c(\mathbf{r}_\mathrm{A})]
 \me^{\mi\Omega(\mathbf{r}_\mathrm{A})(t-t_0)}
 \!-\!\sin^2[\theta_c(\mathbf{r}_\mathrm{A})]
 \me^{-\mi\Omega(\mathbf{r}_\mathrm{A})(t-t_0)}\!\Bigr\}\!\Bigr)\\
+\mathrm{C.c.}
\end{multline}
Here, we have used the identities
\begin{align}
\label{eq100}
&\frac{\Omega_\mathrm{R}(\mathbf{r}_\mathrm{A})}
 {\Delta(\mathbf{r}_\mathrm{A})
 -\Omega(\mathbf{r}_\mathrm{A})}
=-\tan\theta_c(\mathbf{r}_\mathrm{A}),\\[1ex]
\label{align}
&\frac{\Omega_\mathrm{R}(\mathbf{r}_\mathrm{A})}
 {\Delta(\mathbf{r}_\mathrm{A})
 +\Omega(\mathbf{r}_\mathrm{A})}
=\cot\theta_c(\mathbf{r}_\mathrm{A}),
\end{align}
which follow from Eqs.~(\ref{eq88b}) and (\ref{eq89}) upon applying
Eq.~(\ref{eq41.1}). Using Eq.~(\ref{eq95x}) for real dipole matrix
elements, recalling Eq.~(\ref{eq39.1}) and employing the
relation
\begin{gather}
\label{eq103.1}
\bm{\nabla}\sqrt{\Omega_\mathrm{R}^2(\mathbf{r})
 +\Delta^2(\mathbf{r}_\mathrm{A})}
 \big|_{\mathbf{r}=\mathbf{r}_\mathrm{A}}
 =\sin[2\theta_c(\mathbf{r}_\mathrm{A})]
 \bm{\nabla}_{\!\!\mathrm{A}}
 \Omega_\mathrm{R}(\mathbf{r}_\mathrm{A}),
\end{gather}
as implied by Eqs.~(\ref{eq41.1}), (\ref{eq88b}), and (\ref{eq89}),
we arrive at the final result
\begin{multline}
\label{eq103}
\mathbf{F}_\theta^\mathrm{el}(\mathbf{r}_\mathrm{A},t)
=\me^{-\gamma(\mathbf{r}_\mathrm{A})(t-t_0)}\bigl(
 \cos\bigl\{2[\theta\!-\!\theta_c(\mathbf{r}_\mathrm{A})]\bigr\}\\
 +\cot[2\theta_c(\mathbf{r}_\mathrm{A})]
 \sin\bigl\{2[\theta\!-\!\theta_c(\mathbf{r}_\mathrm{A})]\bigr\}
 \cos[\Omega(\mathbf{r}_\mathrm{A})(t-t_0)]\bigr)\\
 \times\mathbf{F}_+(\mathbf{r}_\mathrm{A}),
\end{multline}
where
\begin{equation}
\label{eq103.2}
\mathbf{F}_+(\mathbf{r}_\mathrm{A})
=-\frac{\hbar}{2}\,\bm{\nabla}\!\sqrt{\Omega_\mathrm{R}^2(\mathbf{r})
 +\Delta^2(\mathbf{r}_\mathrm{A})}
 \big|_{\mathbf{r}=\mathbf{r}_\mathrm{A}}.
\end{equation}

\begin{figure}[!t!]
\begin{center}
\includegraphics[width=\linewidth]{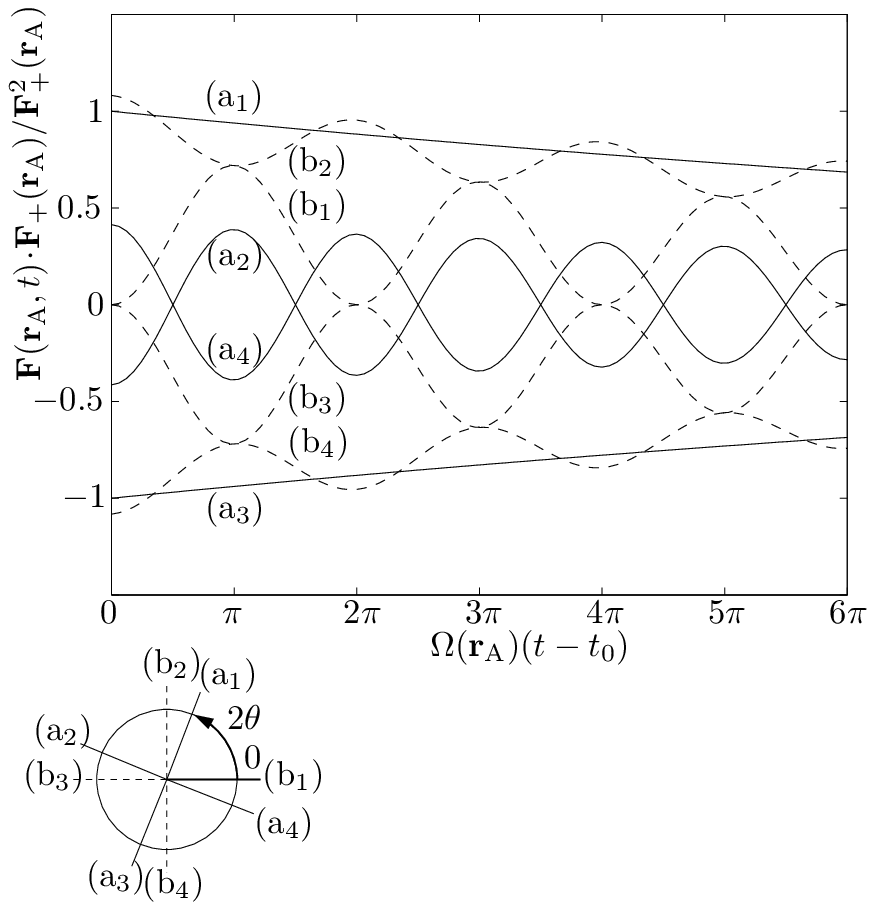}
\end{center}
\caption{
\label{fig2}
The time dependence of the electric part of the resonant CP force for
strong atom--field coupling is displayed for different values of
$\theta$, i.e., $2\theta$ $\!=$
$2\theta_\mathrm{c}(\mathbf{r}_\mathrm{A})$ ($\mathrm{a}_1$),
$2\theta_\mathrm{c}(\mathbf{r}_\mathrm{A})\!+\!\pi/2$
($\mathrm{a}_2$),
$2\theta_\mathrm{c}(\mathbf{r}_\mathrm{A})\!+\!\pi$ ($\mathrm{a}_3$),
$2\theta_\mathrm{c}(\mathbf{r}_\mathrm{A})\!+\!3\pi/2$
($\mathrm{a}_4$),
$\!0$ ($\mathrm{b}_1$), $\pi/2$ ($\mathrm{b}_2$),
$\pi$ ($\mathrm{b}_3$), $3\pi/2$ ($\mathrm{b}_4$), with parameters
$2\theta_\mathrm{c}(\mathbf{r}_\mathrm{A})$ $\!=$ $3\pi/8$ and
$\gamma(\mathbf{r}_\mathrm{A})$ $\!=$
$\!0.05\,\Omega(\mathbf{r}_\mathrm{A})$. The angles $2\theta$ for the
various curves are indicated in the small polar diagram.
}
\end{figure}

We first observe that at initial time $t_0$, Eq.~(\ref{eq103}) reads
\begin{multline}
\label{eq104}
\mathbf{F}_\theta
^\mathrm{el}
(\mathbf{r}_\mathrm{A},t_0)
=-\frac{\hbar}{2}\,\bigl(
 \cos\bigl\{2[\theta\!-\!\theta_c(\mathbf{r}_\mathrm{A})]\bigr\}
 +\cot[2\theta_c(\mathbf{r}_\mathrm{A})]\\
 \times
 \sin\bigl\{2[\theta\!-\!\theta_c(\mathbf{r}_\mathrm{A})]\bigr\}
 \bigr)
 \bm{\nabla}\sqrt{\Omega_\mathrm{R}^2(\mathbf{r})
 +\Delta^2(\mathbf{r}_\mathrm{A})}
 \big|_{\mathbf{r}=\mathbf{r}_\mathrm{A}},
\end{multline}
which agrees with the result (\ref{eq43.8}) [together
with Eq.~(\ref{eq43c})] found in Sec.~\ref{sec3.1},
when neglecting the frequency shift
$\delta\omega'_1(\mathbf{r}_\mathrm{A})$, i.e.,
making the replacement
\mbox{$\Delta(\mathbf{r}_\mathrm{A}) \mapsto\Delta$}.
Equation~(\ref{eq103}) further shows that the electric part of the
force as a function of time is always damped
by an overall exponential
factor and that it contains an oscillating term whose relative
strength depends on both the coupling angle
$\theta_\mathrm{c}(\mathbf{r}_\mathrm{A})$ and the initial state of
the system,~$|\theta\rangle$. This is illustrated in Fig.~\ref{fig2},
where the time dependence of the electric part of the force is
displayed for various initial states $|\theta\rangle$ and fixed
coupling angle $2\theta_\mathrm{c}(\mathbf{r}_\mathrm{A})$ $\!=$
$3\pi/8$. The curves in the figure can be grouped in pairs of curves
differing only by their sign, each of these pairs corresponds to
values of $\theta$ which differ by $2\Delta\theta$ $\!=$ $\!\pi$. The
existence of such pairs is an obvious consequence of
Eq.~(\ref{eq103}). The figure further reveals that there are two
extremes of behavior: While for the initial states with $2\theta$
$\!=$ $\!2\theta_\mathrm{c}(\mathbf{r}_\mathrm{A})$,
$2\theta_\mathrm{c}(\mathbf{r}_\mathrm{A})\!+\!\pi$, the force
shows no oscillations and is purely exponentially damped as a function
of time [curves ($\mathrm{a}_1$) and ($\mathrm{a}_3$)], the initial
states $2\theta$ $\!=$
$2\theta_\mathrm{c}(\mathbf{r}_\mathrm{A})\!+\!\pi/2$,
$2\theta_\mathrm{c}(\mathbf{r}_\mathrm{A})\!+\!3\pi/2$ lead to
oscillations of maximal amplitude around zero [curves ($\mathrm{a}_2$)
and ($\mathrm{a}_4$)]. For other values of $\theta$, the temporal
behavior of the force is a superposition of oscillating and
non-oscillating components [curves
($\mathrm{b}_1$)--($\mathrm{b}_4$)].

In the case $2\theta$ $\!=$ $\!0$, which corresponds to the initial
state $|1\rangle|\{0\}\rangle$, the oscillating and nonoscillating
contributions to the force combine in such a way that the sign of the
force remains unchanged for all times [curve ($\mathrm{b}_1$)]---in
agreement with Eq.~(\ref{eq95.5}). The same is valid for the force in
the case of the initial state $|0\rangle|1_\nu\rangle$
\mbox{($2\theta$ $\!=$ $\!\pi$)} which has just the opposite global
sign [curve ($\mathrm{b}_3$)]. At a first glance, one might expect
that the respective states evolve into each another during the course
of time so that the different global signs would be contradictory.
To resolve the apparent contradiction, one must ask whether the
initial state $|1\rangle|\{0\}\rangle$ ever evolves into the state
$|0\rangle|1_\nu\rangle$. A simple calculation reveals that the answer
is never: Combining Eqs.~(\ref{eq58}), (\ref{eq67}), (\ref{eq87}),
(\ref{eq90}), and (\ref{eq91}) for $2\theta$ $\!=$ $\!0$, we find that
for the state $|\psi(t)\rangle$ that is initially prepared in the
state $|1\rangle|\{0\}\rangle$ the probability $|\psi_1(t)|^2$ reads
\begin{multline}
\label{eq104b}
|\psi_1(t)|^2=\cos^4[\theta_c(\mathbf{r}_\mathrm{A})]
 +\sin^4[\theta_c(\mathbf{r}_\mathrm{A})]
 +2\cos^2[\theta_c(\mathbf{r}_\mathrm{A})]\\
 \times\sin^2[\theta_c(\mathbf{r}_\mathrm{A})]
 \cos[\Omega(\mathbf{r}_\mathrm{A})(t-t_0)]\\
 \ge\cos^2[2\theta_c(\mathbf{r}_\mathrm{A})],
\end{multline}
showing that the state $|0\rangle|1_\nu\rangle$ is indeed never
reached unless in the case of exact resonance,
$\Delta(\mathbf{r}_\mathrm{A})$ $\!=$ $\!0$
[$\theta_c(\mathbf{r}_\mathrm{A})$ $\!=$ $\!\pi/4$], where the
(resonant parts of the) forces associated with both initial states
identically vanish.

Using the same approximations leading from Eq.~(\ref{eq95.2}) to
Eq.~(\ref{eq103}), we find that in the limit of strong atom--field
coupling, the magnetic component of the force, Eq.~(\ref{F4}),
can be given as
 \begin{multline}
\label{4.121x}
\vect{F}_\theta^\mathrm{mag}(\vect{r}_\mathrm{A},t)
=\me^{-\gamma(\vect{r}_\mathrm{A})(t-t_0)}\\
 \frac{\sin\bigl\{2[\theta\!-
\!\theta_\mathrm{c}(\vect{r}_\mathrm{A})]
\bigr\}}
 {\sin[2\theta_\mathrm{c}(\vect{r}_\mathrm{A})]}\,
 \cos[\Omega(\vect{r}_\mathrm{A})(t-t_0)]\\
\times\mu_0\omega_\nu\gamma_\nu
 \vect{d}_{10}\vprod\bigl[
 \vect{\nabla}\vprod\ten{G}^{(1)}
 (\vect{r},\vect{r}_\mathrm{A},\omega_\nu)
 \sprod\vect{d}_{01}
 \bigr]_{\vect{r}=\vect{r}_\mathrm{A}}.
\end{multline}
Comparing this equation with Eq.~(\ref{eq103}) for the electric
component of the force, we see that the magnetic component has a quite
different vector structure than the electric one, and its order of
magnitude is roughly $\Omega(\vect{r}_\mathrm{A})/\omega_\nu$ times
that of the electric component, so that it might become relevant
in the context of the recently considered superstrong coupling regime
\cite{0740}. In particular, the magnetic component of the force
vanishes when the system is initially prepared in the state
$|\theta\!=\!\theta_\mathrm{c}(\vect{r}_\mathrm{A})\rangle$ or the
state
$|\theta\!=\!\theta_\mathrm{c}(\vect{r}_\mathrm{A})\!+\!\pi/2\rangle$
for which the electric component of the force is nonoscillating.
In all the other cases, the magnetic component is---in contrast
to the electric component---always purely oscillating around zero.

Finally, let us compare the nonoscillating force
\begin{equation}
\label{eq105}
\mathbf{F}_\pm(\mathbf{r}_\mathrm{A},t)
=\mp\frac{\hbar}{2}\,\me^{-\gamma(\mathbf{r}_\mathrm{A})(t-t_0)}
 \bm{\nabla}\!\sqrt{\Omega_\mathrm{R}^2(\mathbf{r})
 +\Delta^2(\mathbf{r}_\mathrm{A})}
 \big|_{\mathbf{r}=\mathbf{r}_\mathrm{A}},
\end{equation}
which is observed when the system is initially prepared in an
approximate energy eigenstate, i.e.,
$|\theta\!=\!\theta_\mathrm{c}(\vect{r}_\mathrm{A})\rangle$ or
$|\theta\!=\!\theta_\mathrm{c}(\vect{r}_\mathrm{A})\!
+\!\pi/2\rangle$,
with the corresponding static result [Eq.~(\ref{eq43b})
together with Eqs.~(\ref{eq43c}) and (\ref{eq43})]. We see
that (i) the static approximation may be regarded as being a
good approximation on time scales which are small compared to
$\gamma^{-1}(\mathbf{r}_\mathrm{A})$, and (ii) the
atom--field detuning~(\ref{eq62}), which enters the force, is
different from its bare value (\ref{eq39}) due to the coupling with
the residual field. However, from Eq.~(\ref{eq95}) this effect may be
expected to be small in general.


\section{Summary}
\label{sec4}

Based on macroscopic QED in linear media, we have developed a general
theory of the resonant CP force experienced by a two-level atom
in the presence of arbitrary linear bodies, with special emphasis
on strong atom--field coupling. Assuming that the initial state is a
(coherent) superposition of states that carry a single excitation
quantum each, we have first worked within a static approximation.
Reducing the Hilbert space of the system to an approximately invariant
two-dimensional subspace on which the Hamiltonian assumes a
Jaynes--Cummings form, the eigenenergies and eigenstates have been
constructed according to the well-known dressed-states approach.
Identifying the position-dependent part of the eigenenergies with the
CP potential for the system being prepared in a dressed state, a
simple intuitive picture for the CP force has been obtained,
generalizing results obtained for cavities to arbitrary resonator-like
equipments.

As the static approximation does not take into account the decay of
excited states due to unavoidable radiative and non-radiative losses,
the result found for a system prepared in a dressed state is only
valid on a time scale which is short compared to the time scale of
decay. For systems initially prepared in other than dressed states,
the static approximation is even more problematic, because it also
neglects the Rabi dynamics which is to be expected in the
strong-coupling regime. Motivated by these shortcomings, we have
developed an alternative, dynamical approach to the problem by
starting from the operator-valued Lorentz force and identifying the CP
force with its expectation value where the respective state vector of
the system solves the time-dependent Schr\"{o}dinger equation for
given initial condition. By separating the body-assisted field into
the part that is in (quasi-)resonance with the atomic transition  and
strongly interacts with the atom and the residual part that weakly
interacts with the atom, we have solved the Schr\"{o}dinger equation
in rotating wave approximation to get a solution which fully
incorporates the dynamics induced by both parts of the field. As a
consequence, a general expression for the time-dependent resonant CP
force has been obtained.

For weak atom--field coupling, this expression reproduces the force
obtained earlier for the case where the atom is initially in the upper
state and the field is in the vacuum state. The dynamic behavior of
the force, which at the initial time effectively agrees with the force
obtained in leading-order time-independent perturbation theory, is
given by an exponential damping due to spontaneous decay. For strong
atom--field coupling, different dynamical behaviors are possible,
depending on the initial preparation of the combined system. When the
system is prepared in a dressed state, the force which initially
agrees with the force obtained in static approximation, undergoes an
exponential decay in the further course of time which results from the
width of the upper atomic energy level and the width of the
body-assisted nonmonochromatic mode. When the initial state is a more
general (single excited) state of the atom--field system, then damped
Rabi oscillations of the force are observed, whose amplitude and mean
value sensitively depend on the chosen initial state. In particular,
when the atom is initially excited with the field being in the vacuum
state, the force due to the electric field exhibits two major
differences with respect to the weak-coupling result: It undergoes
damped Rabi oscillations and it is scaled by a correction factor.
Furthermore, it has been found that while the dressed-state force is
entirely due to the interaction of the atom with the body-assisted
electric field, for a more general initial state, additional
oscillating force components appear that result from the interaction
of the atom with the magnetic field. The general results obtained can
be applied to actual geometries by using the appropriate Green
tensors, in order to analyze the respective spatial structure of the
force in more detail.


\acknowledgments

We acknowledge discussions with M. Khanbekyan and C. Raabe.
This work was supported by the Deutsche Forschungsgemeinschaft.


\appendix

\section{Single-resonance approximation}
\label{AppA}

Applying Eq.~(\ref{eq57}), the last term on the r.h.s. of
Eq.~(\ref{eq56}) can be written as
\begin{multline}
\label{A1}
-\int_0^\infty\mathrm{d}\omega\,
 g^2(\mathbf{r}_\mathrm{A},\omega)
 \int_{t_0}^t\mathrm{d}\tau\,
 \me^{-\mi(E_0/\hbar+\omega)(t-\tau)}\psi_1(\tau)\\
=-\int_0^\infty\mathrm{d}\omega\,
 g^{\prime 2}(\mathbf{r}_\mathrm{A},\omega)
 \int_{t_0}^t\mathrm{d}\tau\,
 \me^{-\mi(E_0/\hbar+\omega)(t-\tau)}\psi_1(\tau)\\
-\frac{\gamma_\nu^2}{4}\,g^2(\mathbf{r}_\mathrm{A},\omega_\nu)
 \int_0^\infty\frac{\mathrm{d}\omega}
 {(\omega-\omega_\nu)^2+\gamma_\nu^2/4}\\
\times\int_{t_0}^t\mathrm{d}\tau\,
 \me^{-\mi(E_0/\hbar+\omega)(t-\tau)}\psi_1(\tau),
\end{multline}
where, in accordance with the assumptions of the single resonance
approximation, the first term may be treated within the Markov
approximation. We hence assume that the function
$\me^{\mi\tilde{E}_1(\mathbf{r}_\mathrm{A})t/\hbar}
\psi_1(t)$ is a slowly varying function, so that
\begin{align}
\label{A2}
-&\int_0^\infty\mathrm{d}\omega\,
 g^{\prime 2}(\mathbf{r}_\mathrm{A},\omega)
 \int_{t_0}^t\mathrm{d}\tau\,
 \me^{-\mi(E_0/\hbar+\omega)(t-\tau)}\psi_1(\tau)\nonumber\\
&\simeq-\psi_1(t)\int_0^\infty\mathrm{d}\omega\,
 g^{\prime 2}(\mathbf{r}_\mathrm{A},\omega)
 \int_{t_0}^t\mathrm{d}\tau\,
 \nonumber\\
&\quad\times\me^{-\mi\bigl[\hbar\omega
 -\tilde{E}_1(\mathbf{r}_\mathrm{A})+E_0\bigr](t-\tau)/\hbar}
 \nonumber\\
&\simeq-\psi_1(t)\int_0^\infty\mathrm{d}\omega\,
 g^{\prime 2}(\mathbf{r}_\mathrm{A},\omega)
 \zeta\bigl(\bigl[\tilde{E}_1(\mathbf{r}_\mathrm{A})-E_0\bigr]/\hbar
 -\omega\bigr)
\end{align}
[$\zeta(x)$ $\!=$ $\!\pi\delta(x)$ $\!+$ $\!\mi\,\mathcal{P}/x$].
Using Eqs.~(\ref{eq27}) and (\ref{eq57}) as well as the identity
\begin{multline}
\label{A3}
\mathcal{P}\int_0^\infty
 \frac{\mathrm{d}\omega}
 {\bigl[\tilde{E}_1(\mathbf{r}_\mathrm{A})-E_0\bigr]/\hbar
 -\omega}\,
 \frac{1}{(\omega-\omega_\nu)^2+\gamma_\nu^2/4}\\
=\frac{2\pi}{\gamma_\nu}\,
 \frac{\bigl[\tilde{E}_1(\mathbf{r}_\mathrm{A})-E_0\bigr]/\hbar
 -\omega_\nu}
 {\bigl\{\bigl[\tilde{E}_1(\mathbf{r}_\mathrm{A})-E_0\bigr]/\hbar
 -\omega_\nu\bigr\}^2+\gamma_\nu^2/4}\,,
\end{multline}
which can easily be verified by means of contour-integral techniques
for $\gamma_\nu/2$ $\!\ll$ $\!\omega_\nu$, we find
\begin{multline}
\label{A4}
-\int_0^\infty\mathrm{d}\omega\,
 g^{\prime 2}(\mathbf{r}_\mathrm{A},\omega)
 \int_{t_0}^t\mathrm{d}\tau\,
 \me^{-\mi(E_0/\hbar+\omega)(t-\tau)}\psi_1(\tau)\\
=\Bigl[-\mi\delta\omega'_1(\mathbf{r}_\mathrm{A})
 -{\textstyle\frac{1}{2}}\Gamma'_1(\mathbf{r}_\mathrm{A})\Bigr]
 \psi_1(t),
\end{multline}
where we have introduced Eqs.~(\ref{eq60})--(\ref{eq62}). Substituting
this back into Eq.~(\ref{A1}) and evaluating the second term by means
of Eqs.~(\ref{eq39.1}) and (\ref{eq36.5b}), we arrive at
\begin{multline}
\label{A6}
-\int_0^\infty\mathrm{d}\omega\,
 g^2(\mathbf{r}_\mathrm{A},\omega)
 \int_{t_0}^t\mathrm{d}\tau\,
 \me^{-\mi(E_0/\hbar+\omega)(t-\tau)}\psi_1(\tau)\\
=\Bigl[-\mi\delta\omega'_1(\mathbf{r}_\mathrm{A})
 -{\textstyle\frac{1}{2}}\Gamma'_1(\mathbf{r}_\mathrm{A})\Bigr]
 \psi_1(t)
 -{\textstyle\frac{1}{4}}\Omega_\mathrm{R}^2(\mathbf{r}_\mathrm{A})\\
\times\int_{t_0}^t\mathrm{d}\tau\,
 \me^{[-\mi(E_0/\hbar+\omega_\nu)-\gamma_\nu/2]
 (t-\tau)}\psi_1(\tau),
\end{multline}
so Eq.~(\ref{eq56}) can be written in the form
\begin{multline}
\label{A7}
\dot{\psi}_1(t)
 =\biggl[-\mi\frac{E_1}{\hbar}
 -\mi\delta\omega'_1(\mathbf{r}_\mathrm{A})
 -{\textstyle\frac{1}{2}}
 \Gamma'_1(\mathbf{r}_\mathrm{A})\biggr]\psi_1(t)\\
-\textstyle{\frac{1}{2}}\mi\Omega_\mathrm{R}(\mathbf{r}_\mathrm{A})
 \sin\theta\,\me^{[-\mi(E_0/\hbar+\omega_\nu)-\gamma_\nu/2](t-t_0)}\\
 -{\textstyle\frac{1}{4}}\Omega_\mathrm{R}^2(\mathbf{r}_\mathrm{A})
 \int_{t_0}^t\mathrm{d}\tau\,
 \me^{[-\mi(E_0/\hbar+\omega_\nu)-\gamma_\nu/2]
 (t-\tau)}\psi_1(\tau).
\end{multline}
By using Eq.~(\ref{eq58}), this result is transformed to
\begin{multline}
\label{A8}
\dot{\phi}_1(t)
=-\textstyle{\frac{1}{2}}\mi\Omega_\mathrm{R}(\mathbf{r}_\mathrm{A})
 \sin\theta\,\me^{\{-\mi\Delta(\mathbf{r}_\mathrm{A})-[\gamma_\nu
 -\Gamma'_1(\mathbf{r}_\mathrm{A})]/2\}(t-t_0)}\\
-{\textstyle\frac{1}{4}}\Omega_\mathrm{R}^2(\mathbf{r}_\mathrm{A})
 \int_{t_0}^t\mathrm{d}\tau\,
 \me^{\{-\mi\Delta(\mathbf{r}_\mathrm{A})-[\gamma_\nu
 -\Gamma'_1(\mathbf{r}_\mathrm{A})]/2\}(t-\tau)}
 \phi_1(\tau)
\end{multline}
and after differentiating w.r.t. $t$, we arrive at Eq.~(\ref{eq65})
together with Eq.~(\ref{eq66}).


\end{document}